\documentclass[pdflatex,sn-nature]{sn-jnl}

\usepackage{graphicx}%
\usepackage{multirow}%
\usepackage{amsmath,amssymb,amsfonts}%
\usepackage{amsthm}%
\usepackage{mathrsfs}%
\usepackage[title]{appendix}%
\usepackage{xcolor}%
\usepackage{textcomp}%
\usepackage{manyfoot}%
\usepackage{booktabs}%
\usepackage{algorithm}%
\usepackage{algorithmicx}%
\usepackage{algpseudocode}%
\usepackage{listings}%
\usepackage{graphicx}
\usepackage{subcaption}
\usepackage{bbm}
\usepackage{makecell}
\usepackage{verbatim}
\usepackage{booktabs}
\usepackage{tabularx}
\usepackage{array}
\newcolumntype{L}[1]{>{\raggedright\arraybackslash}p{#1}}
\newcolumntype{Y}{>{\raggedright\arraybackslash}X}
\usepackage{tabularray}
\usepackage{xr}

\theoremstyle{thmstyleone}%
\theoremstyle{thmstyletwo}%

\theoremstyle{thmstylethree}%

\begin{document}

\title[Article Title]{
Counterfactual assessment of social media interventions by separating exposure from engagement
}

\author[1,2]{\fnm{Tommaso} \sur{Pisani}}\email{tommaso.pisani@uniroma1.it}

\author[1,2]{\fnm{Pietro} \sur{Gravino}}\email{pietro.gravino@sony.com}
\equalcont{These authors contributed equally to this work.}

\author*[1,2]{\fnm{Giulio} \sur{Prevedello}}\email{giulio.prevedello@sony.com}
\equalcont{These authors contributed equally to this work.}

\affil*[1]{\orgname{Sony Computer Science Laboratories Paris}, \orgaddress{\street{6 rue Amyot}, \city{Paris}, \postcode{75005}, \country{France}}}


\affil[2]{\orgdiv{Centro Ricerche Enrico Fermi}, \orgaddress{\street{Via Panisperna 89/A}, \city{Rome}, \postcode{00185}, \country{Italy}}}

\abstract{
Social media interventions are usually evaluated in individual-level experiments, which are costly to scale and often miss collective effects.
We propose the Exposure--Engagement model, whose parameters separately control what social media users are exposed to (recommendation) and how they engage (behaviour).
After fitting it to the Italian Twitter debate (2018--2022), we validate it at individual and collective levels.
We simulate behavioural interventions (prebunking and accuracy prompts), calibrated on effect sizes, and a recommender-system intervention (chronological feed), comparing effects on engagement, attention diversity, and segregation.
All three reduce engagement. Behavioural interventions slightly reduce diversity, leaving segregation unchanged, whereas the chronological feed increases diversity and reduces segregation.
For these interventions, changing recommendation does more for the debate's structure than changing behaviour.
We design a recommender-system intervention that increases engagement and diversity while reducing segregation, although with about a tenth of the chronological feed's gains, showing the framework's potential for pro-social intervention design.
}


\maketitle

\section{Introduction}\label{sec:intro}

Social media have become central to public debate~\cite{bakshy2012role, kwak2010twitter, vosoughi2018spread}, and a growing body of work links them to rising polarisation~\cite{kubin2021role, allcott2020welfare} and to the segregation of audiences into echo chambers~\cite{Echo_chamber}, although the evidence varies across outcomes and settings~\cite{lorenz2023systematic}.
The online public debate unfolds as the leaders of public discourse (news media, political organisations, and public figures) express different viewpoints, generating opposing sides~\cite{Polarization, 1fab}. The friction between these sides is exacerbated by algorithmic recommendation of moral and emotional content~\cite{moral_values}, which amplifies in-group/out-group bias and triggers adversarial behaviour~\cite{brady2017emotion, bail2018exposure, Toxicity, rathje2021out, marino2024mis}, while the interplay between information supply, public demand, and content diffusion shapes the prevalence and reach of disinformation~\cite{wardle2017information, pennycook2021psychology, Disinformation, gravino2022supply,lazer2018science}.

To reduce polarisation and tackle disinformation, the scientific community has proposed a range of interventions targeting both user behaviour and platform design~\cite{real_intervention_cohen, pennycook_accuracy, chronological_intervention, kozyreva2024toolbox, windisch2022online, voelkel2024megastudy}. Evaluating them, however, faces three limitations. First, interventions are hard and costly to implement and evaluate at scale~\cite{salganik2019bit}. Second, large--scale deployment makes it difficult to isolate the effect of an intervention from concurrent external causes, given the complexity of these platforms and the exogenous factors that continuously perturb them~\cite{lazer2020computational, lorenz2023systematic, hartmann2025systematic, holliday2025depolarization}.
Third, experiments typically measure outcomes at the level of individual users, although network structure can shape collective outcomes well beyond individual responses~\cite{centola2010spread}. Switching Facebook and Instagram users to a chronological feed, for instance, reduced their engagement without detectable effects on issue and affective polarisation~\cite{guess2023social}, but could not reveal whether the organisation of the debate changed.
In silico models offer a way to explore interventions, from opinion-dynamics models describing how individual preferences, social interactions and external influences~\cite{Loreto_opiniondynamics, carletti2006make, crokidakis2012effects, vaz2010mass} shape collective states~\cite{rainer2002opinion,deffuant2000mixing, axelrod1997dissemination, degroot1974reaching,galam2008sociophysics,sznajd2000opinion, castellano2009statistical} to recent ``AI agent societies'' of large language models (LLMs), in which exogenous effects are controlled by construction~\cite{argyle2023out, park2023generative, Giordano}. LLMs can predict the treatment effects of survey experiments about as accurately as human forecasters, although they overestimate their size~\cite{ashokkumar2026large}. Validating LLM agent societies against observed collective behaviour, however, remains a central open challenge~\cite{larooij2026validation}.

Most existing work modelling individual-level user responses on social media has studied exposure, for instance to ideologically diverse news \cite{bakshy2015exposure}, engagement with misinformation \cite{fake_news_intervention}, or both jointly \cite{del2016spreading}, on Facebook. Closer to our work, Bak-Coleman et al.~\cite{bak2022combining} applied epidemic spreading models to simulate the systemic impact of combined interventions against misinformation on Twitter; Guess et al.~\cite{guess2023reshares} ran a field experiment on Facebook that removed reshared content from users' feeds, measuring individual-level outcomes including exposure, engagement, news knowledge and polarisation. However, both studies remain focused on the political debate and misinformation rather than on the broader public debate. Moreover, neither captures the interplay between algorithmic curation and user behaviour that shapes the collective organisation of the system: Guess et al.~\cite{guess2023reshares} measured outcomes at the individual level, while Bak-Coleman et al.~\cite{bak2022combining} tracked aggregate engagement volumes in a model without network structure or content curation, a limitation the authors themselves note, so that intervention coverage could not interact with the relational fabric of the debate.

Here, we address these gaps by providing a framework to assess the systemic impact of interventions on the active core of a social debate, before large-scale deployment. We propose the Exposure--Engagement model, a parametric model which disentangles two fundamental mechanisms: the \textit{exposure} stage, which models what content each user sees as a result of the platform's recommendation, and the \textit{engagement} stage, which models whether users interact with the content they are shown.
We fit the model parameters to the Italian public debate on Twitter (now X) from 2018 to 2022, considering retweets as the interaction of interest. 
The fitted parameters provide an interpretable description of the national debate's month-to-month dynamics, conditioned on the content posted by its leaders. Once the model descriptive adequacy is established, at both the individual user level and the collective system level, interventions can be mapped onto the specific mechanisms they target, behavioural or algorithmic, and simulated as modifications of the corresponding parameters. 
This produces counterfactual scenarios that are otherwise inaccessible in real-world settings and that reveal effects on the collective organisation of the debate, such as its segregation, beyond the reach of individual experiments. Each intervention is also applied to a randomly drawn fraction of users, allowing us to observe how intervention effects scale with population coverage.

Within this framework, we simulate interventions targeting both user behaviour and platform recommendation, evaluating their impact on three system-level properties: user engagement, attention diversity, and segregation. For behavioural interventions, we translate empirically measured effect sizes from prebunking~\cite{real_intervention_cohen} and accuracy prompt~\cite{pennycook_accuracy} studies into changes in the engagement parameters. For the algorithmic intervention, we simulate the transition from a platform-curated to a chronological feed~\cite{chronological_intervention}, a scenario of direct relevance to the European Digital Services Act~\cite{DSA2022}, which requires very large platforms to offer at least one recommender option not based on profiling.

We find that the behavioural interventions produce modest reductions in engagement and attention diversity and leave segregation essentially unchanged, consistent with their small experimental effect sizes. The chronological feed, by contrast, increases attention diversity and reduces segregation in every month when fully deployed, at a similarly modest cost in engagement: a system-level effect that individual-level experiments, which found no change in attitudes, are not designed to capture. The designed intervention improves all three properties at once, but its gains in diversity and segregation are about one tenth of those of the chronological feed, suggesting a trade-off between engagement and pro-social outcomes within the Exposure--Engagement model.

Because all interventions are evaluated on the same data and the same three properties, their system-level effects can be compared directly and tracked as coverage increases, even though they were originally tested in different settings. For the interventions studied here, this comparison suggests that changing what users see does more for the collective structure of the debate than changing how they respond to it.

Finally, we invert the assessment procedure: guided by an exploratory analysis of the parameter space, we design a novel combined intervention that simultaneously increases user engagement and attention diversity while reducing segregation, that is the desired outcome for a sustainable, pro-social intervention. This illustrates the framework's potential as a tool not only for counterfactual assessment, but also for intervention design.

\section{Results}\label{sec:results}
\subsection{Modelling the online public debate}

The purpose of our framework is to simulate the behaviour of a complex system composed of social media users and the algorithm that influences their interactions. By explicitly separating these two aspects, we gain control and interpretability, and we can simulate the outcomes of interventions on specific parts of the behaviour or the algorithm, as well as any possible combination. In this work, we focus on the public debate that unfolded on Twitter in Italy between 2018 and 2022, by monitoring the most prominent politicians' and news outlets' accounts and tweets, and using other users' retweets as a proxy for engagement (see Methods). The same approach can potentially be applied to other kinds of accounts (e.g. influencers), other platforms and other kinds of engagement. The dynamics will be considered aggregated on ``timeframes'' $t$. While the model supports other timeframe sizes, in this work, we aggregate monthly to capture a sufficient signal while avoiding over-aggregation of dynamics.

\begin{figure}[htbp]
    \centering
        \centering
        \includegraphics[width=\textwidth, trim={2cm 0 2cm 0},clip]{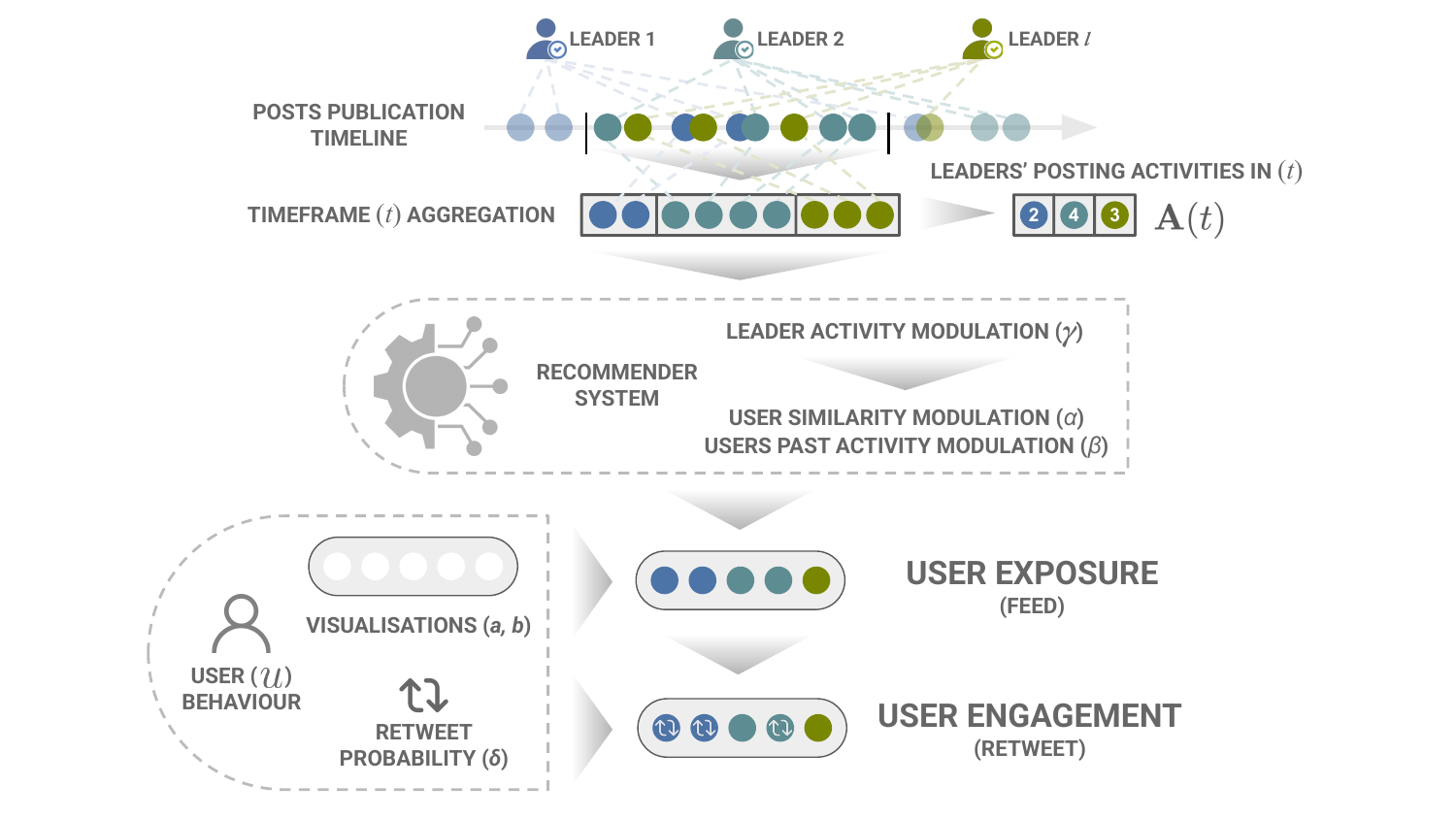}
        \caption{A graphic representation of the Exposure--Engagement model. From the leaders' activity, the model simulates the recommender system and the user behaviour, controlling them with six interpretable parameters.}
    \label{fig:model_image}
\end{figure}

\subsubsection{The Exposure--Engagement model}
The system we are modelling comprises the platform's algorithm, which governs content exposure through a recommender system, and the users' accounts, which produce and interact with content. The accounts are divided into $M$ ``leaders'', i.e. influential accounts of news outlets and politicians, and $N$ ``followers'' or ``users'', i.e. all the other accounts interacting with the leaders (see Methods for more details).

\paragraph{The main observables: leaders' activities and the retweet matrix}
The dynamics we observe in the data in each timeframe are followers' engagement reactions (i.e. retweets) to leaders' posting activities. The latter is represented by the $M$-dimensional vector $\mathbf{A}(t)$, whose generic element $A_l(t)$ is the total number of posts published by leader $l$ in timeframe $t$. On the other hand, the $N \times M$ retweet matrix $\mathbf{R}(t)$ represents the users' reactions. Its generic element $R_{ul}(t)$ is the number of times user $u$ retweeted content from leader $l$ during timeframe $t$. This matrix is crucial because it contains the engagement patterns used to assess the system's status at the individual level (e.g., user's engagement) and at the system level (e.g., segregation).

Our purpose is to simulate $\mathbf{R}(t+1)$ if a given intervention is applied to the system, thereby generating a counterfactual scenario to study the intervention's impact. To do so, we first model how the reactions $\mathbf{R}(t)$ arise from $\mathbf{A}(t)$ (due to, for example, the relations between users and leaders, or the role of the recommender system), and we then apply the model to the following timeframe's activities, $\mathbf{A}(t+1)$. In other words, the leaders' activities are an input to the model, whose output is therefore conditioned on the observed content supply: the model describes the users' response to that supply, rather than forecasting the system's autonomous evolution. Crucially, the model introduces six parameters to govern the system (Fig.~\ref{fig:model_image}) and, as we will see later, to simulate the interventions (Methods).

\paragraph{The model parameters}
The exposure stage explicitly models users' visualisations of leaders' content. For a given timeframe $t$, $\nu_u(t)$ represents the total amount of visualisations for user $u$. We estimate it as a linear function of the user's previous retweeting activity, with two control parameters $a$ and $b$. Then $\nu_u(t)$ is allocated among the leaders' content according to a multinomial-like process shaped by two main factors:
\begin{itemize}
    \item Recommender system: the algorithm tends to expose user $u$ to content retweeted by similar users, in proportion to how strongly those peers concentrate their attention on each leader. The similarity with other users is modulated by the control parameter $\alpha$, while the weight of the peers' past preference shares is regulated by $\beta$.
    \item Leaders' activity: the more frequently leaders post, the higher their chances of being viewed. The impact of the activity level is governed by the parameter $\gamma$.
\end{itemize}

Once the exposure is established, we move on to the engagement stage. We assume each user $u$ has a relation with each leader $l$, which determines the probability that when $u$ visualises a tweet from $l$, they retweet it. We assume this probability remains constant between a timeframe and the following one, so we can estimate the relations for $t$ from the behaviour observed in $t-1$, using a control parameter $\delta$ (see Table~\ref{tab:params} for the role and limit behaviour of each parameter).

For each monthly transition, the six parameters were estimated by minimising the mean absolute logarithmic error between the empirical and the expected retweet matrices (Methods~\ref{sec:Parameter_Optimization}). This measure emphasises relative rather than absolute errors, thus preventing the few most active user--leader pairs from dominating the estimation and letting the model capture the behaviour of users across a larger activity spectrum.

\begin{table}[t]
\centering
\caption{The parameters controlling the Exposure--Engagement model, their role, and their limit behaviour.}
\label{tab:params}
\begin{tabularx}{\linewidth}{
    >{\centering\arraybackslash}p{0.06\linewidth}
    >{\raggedright\arraybackslash}p{0.13\linewidth}
    >{\raggedright\arraybackslash}p{0.25\linewidth}
    >{\raggedright\arraybackslash}X
    >{\raggedright\arraybackslash}X
}
\toprule
Symbol & Name & Role & Param. $\!\to\!0$ & Param. $\!\to\!\infty$ \\
\midrule
$a$      & Effective view-to-retweet ratio
         & Scale factor between tweets viewed and past retweet activity
         & Views collapse to the baseline $b$.
         & Views dominated by past activity. \\
\midrule
$b$      & Baseline viewing
         & Default number of posts viewed by all users
         & Views entirely activity-driven.
         & Views uniform across users, at the baseline level. \\
\midrule
$\alpha$ & User selectivity
         & Influence of similar users on exposure
         & Global homogenisation.
         & Pure personal preference. \\
\midrule
$\beta$  & Preference reinforcement
         & Influence of peers' past preference shares on exposure
         & Peers' preferences flattened across leaders.
         & Only each peer's top leader contributes. \\
\midrule
$\gamma$ & Leader influence
         & Modulation of leaders' posting activity on exposure
         & Posting activity has no effect on exposure.
         & Attention monopolised by the most active leader. \\
\midrule
$\delta$ & Inertia
         & Past behaviour impact on retweet probability
         & Users retweet every exposed post from previously retweeted leaders.
         & Engagement reduced and concentrated on favourite leaders. \\
\bottomrule
\end{tabularx}
\end{table}

\subsubsection{Descriptive adequacy at the user level}\label{res:descriptive_adeq}
We first assessed whether the Exposure--Engagement model captures the systematic month-to-month reorganisation of engagement, beyond what the mere persistence of past behaviour would imply. To this end, we compared the fitted model with a purely inertial baseline that reproduces the retweet matrix $\mathbf{R}(t+1)$ by replicating the matrix observed in the previous month, $\mathbf{R}(t)$. A fitted model that failed to improve on this zero-parameter baseline would indicate that its mechanisms add no descriptive value over pure inertia.

\begin{figure}[htbp]
    \centering
        \centering
        \includegraphics[width=\textwidth]{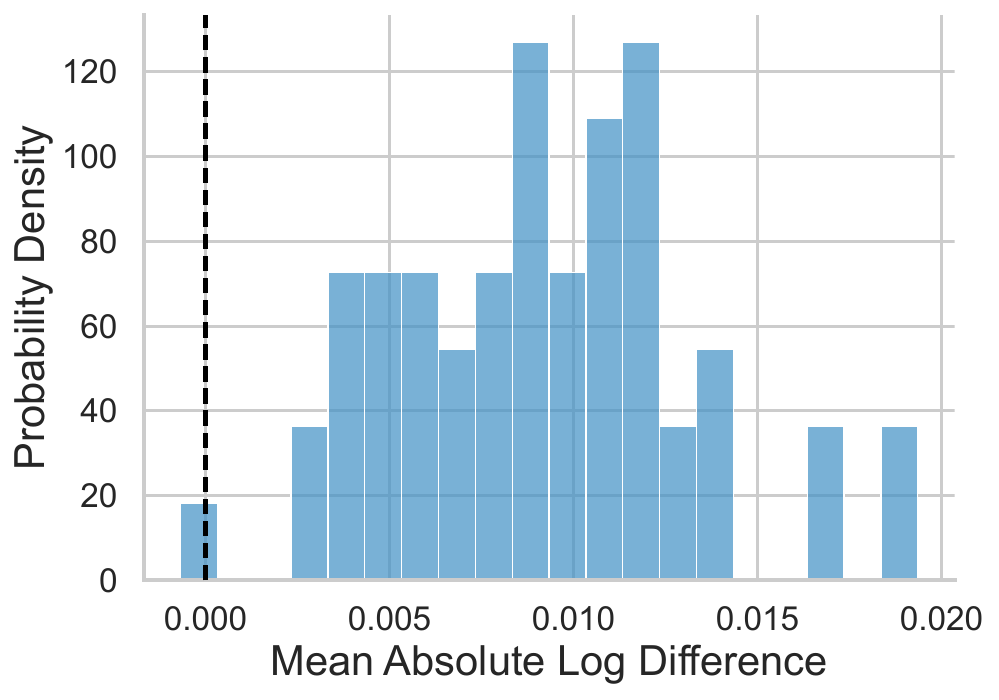}
        \caption{Histogram of all the differences of the mean absolute logarithmic error between our model and the baseline, from 2018 to 2022. The only negative point corresponds to March 2020, a month affected by the exogenous shock from the COVID-19 outbreak in Italy.}
    \label{fig:Baseline_Comparison}
\end{figure}

The model attains a lower discrepancy than the inertial baseline in all months except March 2020 (Figure~\ref{fig:Baseline_Comparison}). Notably, this month coincides with the COVID-19 outbreak in Italy, suggesting that extreme exogenous events can substantially disrupt the system's underlying dynamics, and therefore the model's ability to describe one month from the preceding one; conversely, the consistent improvement in all other months supports the assumption that, at the monthly timescale, the system evolves slowly enough for its dynamics to be captured by the model's mechanisms.

\begin{figure}[htbp]
\centering
\def\pw{0.32\textwidth}

\includegraphics[width=\pw]{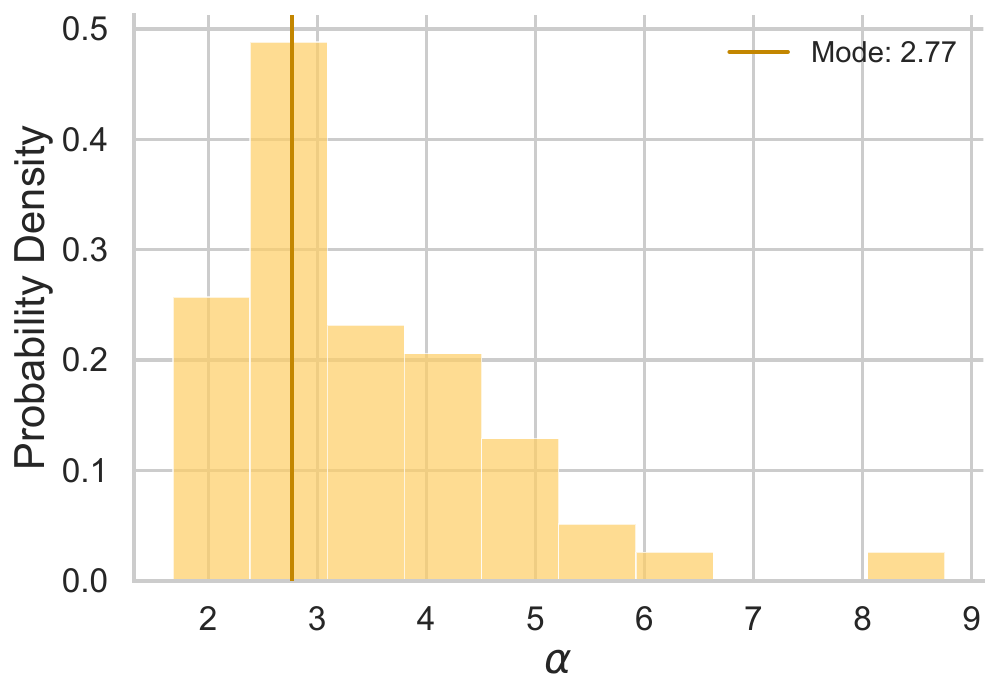}\hfill
\includegraphics[width=\pw]{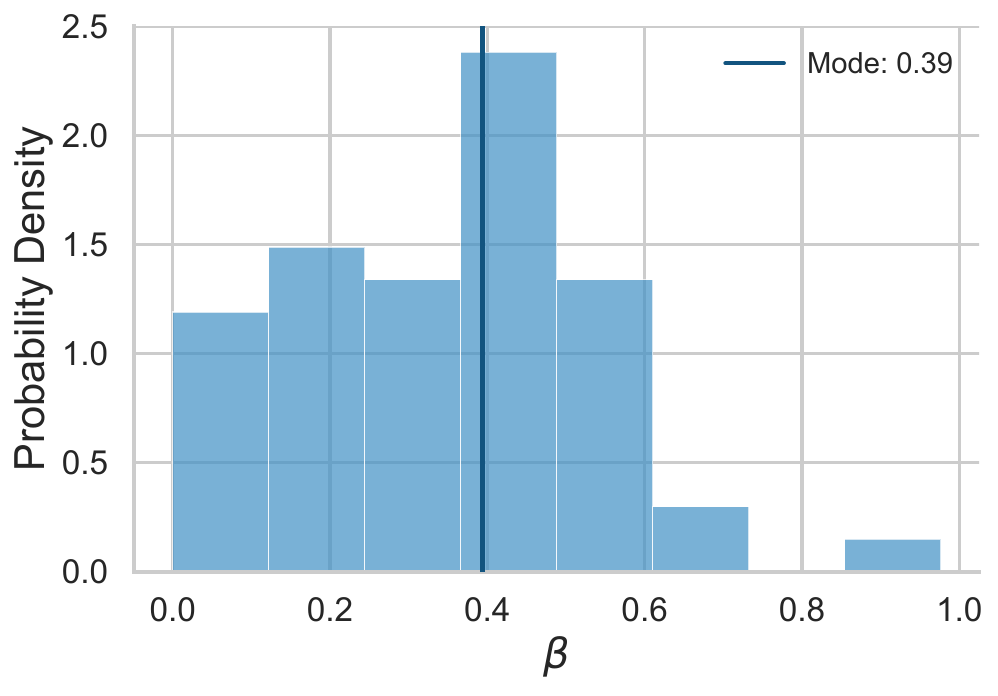}\hfill
\includegraphics[width=\pw]{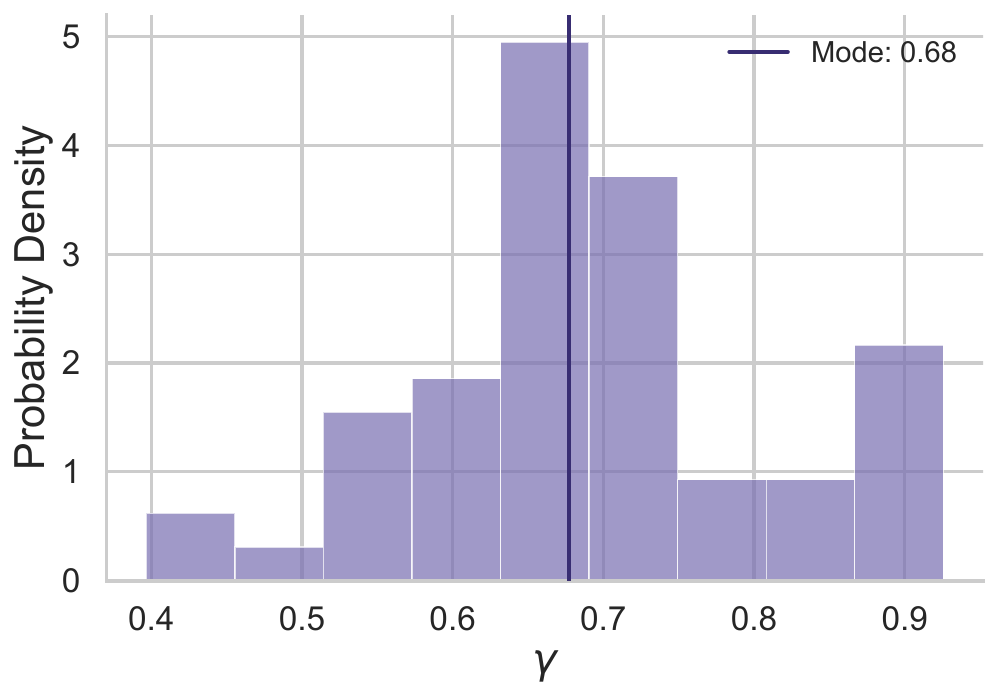}

\vspace{1ex}
\includegraphics[width=\pw]{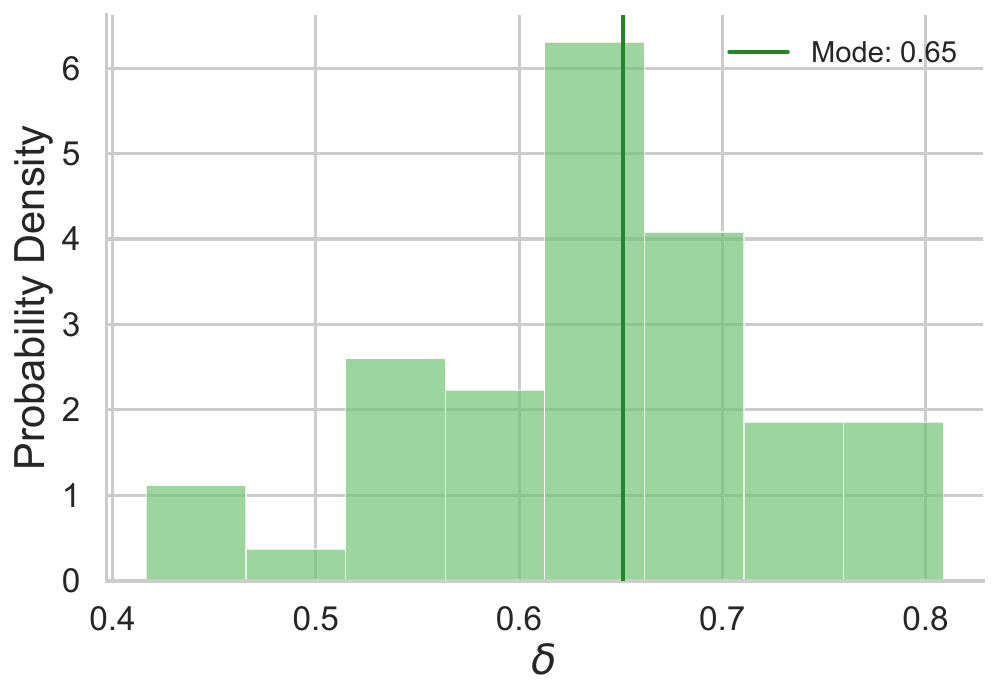}\hfill
\includegraphics[width=\pw]{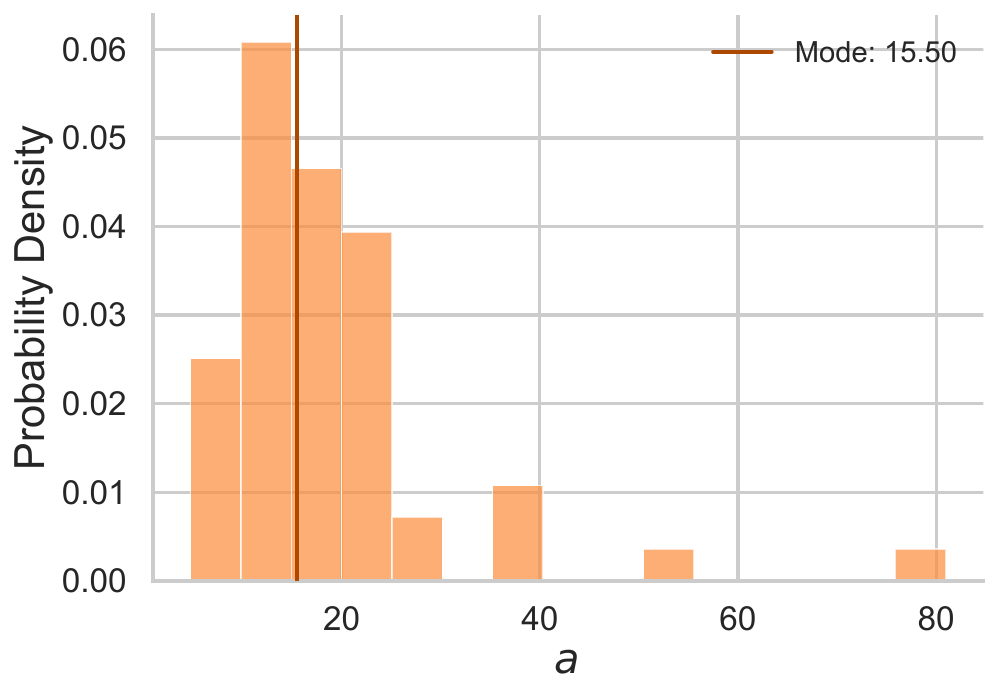}\hfill
\includegraphics[width=\pw]{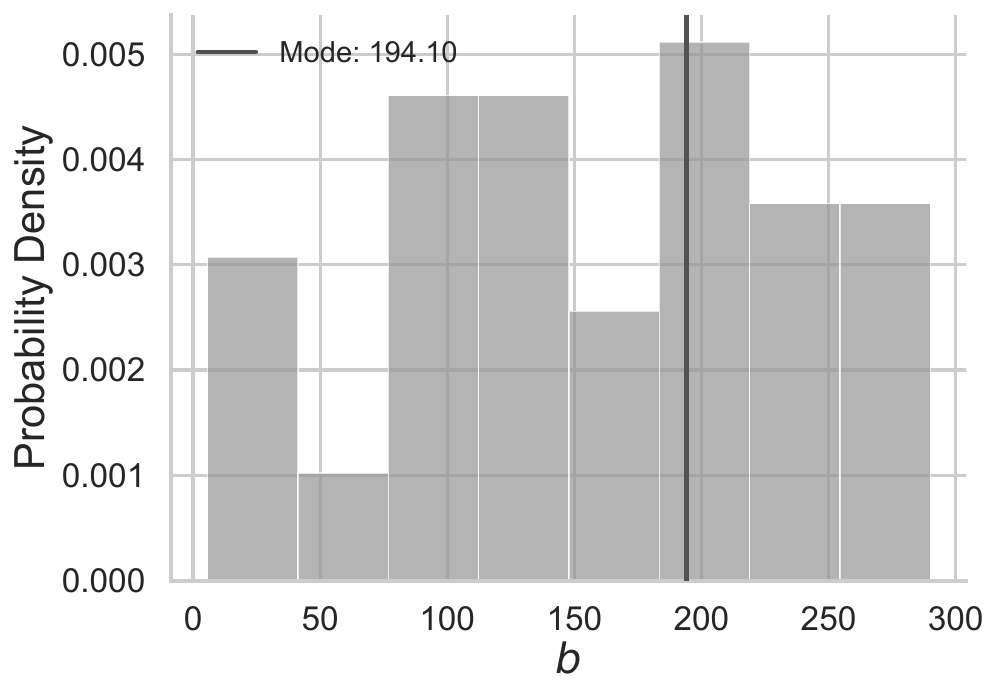}

\caption{Distributions of the fitted parameters over all 55 monthly transitions (February to December, 2018--2022). Top row, from left to right: user selectivity $\alpha$, preference reinforcement $\beta$, leader influence $\gamma$. Bottom row: inertia $\delta$, effective view-to-retweet ratio $a$, baseline viewing $b$. Solid vertical lines mark the modes.}
\label{fig:parameters_distribution}
\end{figure}

To qualitatively assess the temporal stability of the model, we examined the distribution of the six parameters fitted for each month over the five-year period under study. All parameters except $b$, the baseline viewing, show unimodal distributions, with well-defined modes and limited variability (Figure~\ref{fig:parameters_distribution}), indicating that the model exhibits stable behaviour over time. The baseline viewing parameter $b$ shows larger variability, but still in the same order of magnitude. This variability can be explained by the fact that this parameter represents an average of a user's viewing activity distribution. Such a distribution is expected to be fat-tailed, similarly to the retweet activity distribution~\cite{bild2015aggregate}, and therefore, with high variance.

The fitted values of the other parameters are also informative within the model. Each parameter sets how strongly a quantity (user similarity, peers' preference shares, posting activity and past retweet rates, respectively) shapes exposure or engagement: at $1$ the quantity enters proportionally, above $1$ its differences are accentuated, and below $1$ they are attenuated (Methods). Throughout the period, $\alpha > 1$, so exposure depends mostly on the user's own history and on their most similar peers. Conversely, $\beta < 1$, so the recommender spreads exposure across the leaders that similar users engage with, rather than concentrating it on their favourites. Similarly, $\gamma < 1$ means that visibility grows with leaders' posting activity, but less than proportionally, and $\delta < 1$ means that users spread their retweets more evenly across the leaders they already engage with than their past retweets alone would imply.

\subsection{Reproducing collective dynamics}\label{res:collective_dyn}
Having established that the Exposure--Engagement model provides an adequate description of engagement at the level of individual users, we next assessed whether it also reproduces the collective dynamics of the debate.
We focused on three system-level properties relevant for evaluating interventions: user engagement, attention diversity, and segregation. Beyond this validation, we use the same properties in the next section to assess the interventions simulated with the model (Section~\ref{res:Intervention_Assessment}). Because all interventions are simulated on the same data (the Italian debate on Twitter) and evaluated on these same properties, they can be compared directly, even though they were originally tested in different settings and with different outcome measures.

User engagement is measured as the average number of retweets per user in a month. Engagement is central to platforms' business models, so any reduction caused by an intervention is a cost that weighs on its adoption.

Attention diversity is quantified by the average normalised Effective Number of Leaders (nENL; Methods), the fraction of leaders that effectively capture a user's attention. The metric ranges from $1/M$ when retweets concentrate on a single leader to $1$ when retweets are evenly distributed among all leaders. It therefore addresses concerns about filter bubbles and algorithmic amplification, as it detects whether users' attention narrows onto a few sources.

To estimate segregation, we first project the bipartite follower--leader network onto the leader layer and then we measure the modularity of the monopartite leader similarity network (see Methods). High modularity indicates the presence of leader communities that share many followers internally, with little audience overlap across groups, a structural signature of echo-chamber-like fragmentation.
As a property of the network as a whole, modularity captures collective changes in how audiences are shared across the debate, which individual-level experiments cannot observe.

Because the Exposure--Engagement model is stochastic, the model fitted to a given month defines a probability distribution over retweet matrices rather than a single matrix. We therefore drew $10{,}000$ retweet matrices for each month from the fitted model, each conditioned on the empirical matrix of the previous month and on the leaders' observed activity (Methods), and computed the three metrics on each of them, as on the empirical matrix of that month. For each month, we summarise the resulting Monte Carlo distribution of each metric by its mean and its 95\% probability interval (Fig.~\ref{fig:user_metrics_correlations}); these intervals capture only the stochasticity of the model, not the uncertainty from parameter estimation or structural model error.

All three metrics show positive and significant Pearson's correlations between empirical values and simulated means, also after accounting for serial dependence between months (Methods): strong for user engagement ($r = 0.86$) and attention diversity ($r = 0.85$), moderate for segregation ($r = 0.61$; Table~\ref{tab:correlation_metrics_updated}).
Given these linear associations, we described the relationship between simulated and empirical values with a linear fit (Fig.~\ref{fig:user_metrics_correlations} and Table~\ref{tab:correlation_metrics_updated}). For user engagement and segregation, neither the slope nor the intercept differs significantly from perfect agreement (slope $1$, intercept $0$), so the fit is statistically consistent with a one-to-one correspondence between simulated and empirical values. For attention diversity, the slope is significantly below $1$ ($0.56$; $95\%$ CI $0.43$--$0.68$), while the intercept is compatible with zero, indicating that simulated values are compressed to about half of their empirical counterparts. Nevertheless, almost all of the simulated means lie below the identity line for user engagement and attention diversity, and above it for segregation. The model therefore systematically underestimates user engagement and attention diversity and overestimates segregation; for user engagement and segregation, this offset is small relative to the uncertainty of the fit and is therefore not detected by its parameters.

Overall, these biases correspond to a simulated system that appears more segregated, with lower user engagement concentrated on fewer leaders within the same group. A plausible structural origin of these biases lies in the engagement stage, which, by construction, assigns zero retweet probability to any user--leader pair with no interaction in the previous month (see Methods): simulated matrices can therefore lose but never gain user--leader relations, reducing engagement, concentrating each user's attention and sharpening community boundaries (see Discussion).

Despite these offsets, simulated values increase linearly with empirical ones, so a difference between two months has, on average, the same sign in simulation as in the data. Assuming that differences between simulated conditions scale in the same way, they translate to the empirical scale roughly one-to-one for user engagement and segregation (with wide uncertainty for segregation), whereas for attention diversity they are about half the corresponding empirical differences: an empirical difference of $0.01$ nENL corresponds, on average, to $\sim\!0.006$ nENL in simulation. The effects on attention diversity reported below are therefore likely understated. By contrast, the simulated values themselves inherit the offsets described above: the level a metric reaches in a simulated scenario should not be read as the level the real debate would reach, only its difference from the simulated baseline. The following sections thus compare simulated conditions with each other, reporting how consistently and how strongly each intervention changes the three properties.

\begin{figure}[htbp]
    \centering
    \includegraphics[width=\textwidth]{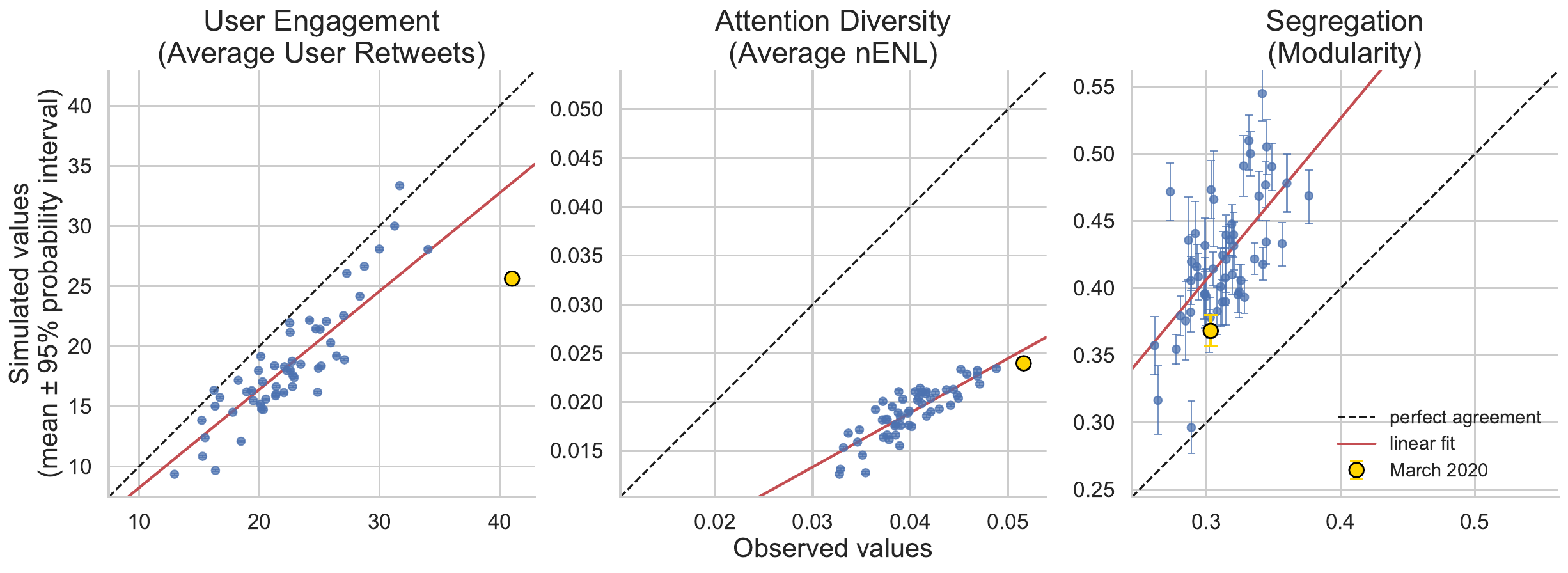}
    \caption{Correlations of system-level metrics between empirical Twitter data and simulated values, from 2018 to 2022: average user retweets for user engagement (left), average nENL for attention diversity (centre), modularity for segregation (right). Each point is one month (February to December, totalling 55 points), with the empirical value on the horizontal axis and the mean of the metric over the $10{,}000$ matrices simulated for that month on the vertical axis. Vertical bars are $95\%$ probability intervals ($2.5$th–$97.5$th percentiles of the simulated values). Intervals narrower than the markers are not visible. The point for March 2020, coinciding with the onset of COVID-19 lockdowns, is highlighted in gold across all three panels. The dashed line is the identity, indicating perfect agreement between empirical and simulated values. The solid red line is the least-squares linear fit of the simulated means on the empirical values (Table~\ref{tab:correlation_metrics_updated}).}
    \label{fig:user_metrics_correlations}
\end{figure}

\begin{table}[hbt!]
\centering
\caption{Correlation statistics and linear relationship parameters between empirical values and simulated means ($n = 55$ months) of the metrics for system-level properties. $p$-values and $95\%$ confidence intervals account for serial dependence between months through the effective sample size (see Methods).}
\label{tab:correlation_metrics_updated}
\begin{tabular}{lrrr}
\toprule
 & \textbf{User engagement} & \textbf{Attention diversity} & \textbf{Segregation} \\
 & Average user engagement & Average nENL & Modularity \\
\midrule
\textbf{Pearson's $r$}       & $0.86$                 & $0.85$                     & $0.61$ \\
\textbf{$p$-value}           & $1.99 \times 10^{-8}$  & $3.89 \times 10^{-10}$     & $2.71 \times 10^{-3}$ \\
\textbf{Slope (95\% CI)}     & $0.82$ $(0.61, 1.02)$  & $0.56$ $(0.43, 0.68)$      & $1.20$ $(0.47, 1.93)$ \\
\textbf{Intercept (95\% CI)} & $0.05$ $(-4.72, 4.82)$ & $-0.003$ $(-0.008, 0.002)$ & $0.05$ $(-0.18, 0.28)$ \\
\bottomrule
\end{tabular}
\vspace{2pt}
\small
\end{table}

\subsection{Intervention assessment and design}\label{res:Intervention_Assessment}
Having established the descriptive adequacy of the Exposure--Engagement model at both the individual and the collective level, we can simulate real-world interventions, scaling their assessment beyond individual experiments to the national debate on social media across several time periods. The framework can then quantify the impact of an intervention on the system's collective dynamics, as measured by the three system-level properties introduced above. To illustrate this, we translate the effects of interventions reported in the literature into changes in the fitted model parameters that preserve the direction and magnitude of the reported effects.

Because the model separates the exposure and engagement stages and its parameters carry interpretable meanings, interventions can be mapped to the specific mechanisms they target. We therefore distinguish \textit{behavioural interventions}, which modify how users interact with content, from \textit{algorithmic interventions}, which modify how the platform ranks and distributes it. Each intervention is applied to a randomly drawn fraction $c \in [0,1]$ of the users (see Methods), and for each intervention we report the resulting change in the three properties relative to the baseline (the fitted model without intervention, $c = 0$) as coverage increases.
To assess the role of deployment scale, we simulated six coverage levels ($c = 0.1$, $0.25$, $0.5$, $0.75$, $0.9$ and $1$) and obtained the distribution of each change by Monte Carlo sampling of both conditions, baseline and $c>0$, in every month.
We summarise each distribution by its mean and by the probability of improvement ($\mathrm{PoI}$), the share of Monte Carlo realisations (simulated retweet matrices) in which the intervention moves a metric in the favourable direction (higher user engagement and attention diversity, lower segregation).

Finally, we demonstrate how the model can inform the design of novel interventions. Rather than mapping a known intervention to parameter changes and observing the outcome, we invert the procedure: we identify parameter configurations that improve global metrics, suggesting candidate interventions targeting the underlying mechanisms.

\subsubsection{Behavioural interventions} \label{sec:results_behavioural_interventions}
We assess two behavioural interventions targeting misinformation sharing, both of which operate through the engagement stage.

\paragraph{Prebunking} Roozenbeek et al.~\cite{real_intervention_cohen} conducted a large-scale field experiment on YouTube in which short videos were shown to users to inoculate them against common manipulation techniques associated with misinformation, such as emotional language and false dichotomies. Effect sizes were measured using Cohen's $h$, a standardised measure for differences in proportions; the strongest effect was observed for the false-dichotomy intervention, with an average Cohen's $h$ of $0.13$.

We translate this effect into the model by mapping each inertia parameter fitted to empirical data, $\delta^{\text{OPT}}$, to a modified value, $\delta^{\text{INT}}$, that reflects the reported effect size. This adjustment applies exclusively to content produced by non-trustworthy sources associated with misinformation (see Methods). This translation rests on several assumptions: that the effect transfers from YouTube to Twitter; that inoculation persists over the full one-month time period used in the model; that reduced misinformation susceptibility translates proportionally to reduced sharing; and that the intervention does not affect user total visualisations, since no evidence on this was reported in the original study.

As shown in Fig.~\ref{fig:intervention_assessment}a, the intervention at full coverage produces a modest reduction in user engagement ($-1.6$ retweets per user) and attention diversity ($-0.001$ nENL) in virtually all realisations ($\mathrm{PoI}=0\%$ and $<0.01\%$, respectively), with both the size and the heterogeneity of these effects growing with coverage (s.d. at full coverage: $0.73$ retweets per user and $0.0004$ nENL). Meanwhile, segregation is essentially unaffected: it decreases in $46\%$ of realisations ($\mathrm{PoI}=46\%$), barely below the $50\%$ expected without an effect, and its mean change is minimal ($-0.001$ modularity). The magnitude of these effects is consistent with the small effect size reported in the original study. By construction, the intervention lowers retweet probabilities for non-trustworthy sources without changing the number of posts viewed, which directly reduces user engagement and concentrates each user's retweets on fewer leaders, lowering attention diversity.

\begin{figure}[p]
    \centering
    \begin{subfigure}[t]{0.9\textwidth}
        \centering
        \includegraphics[width=\textwidth]{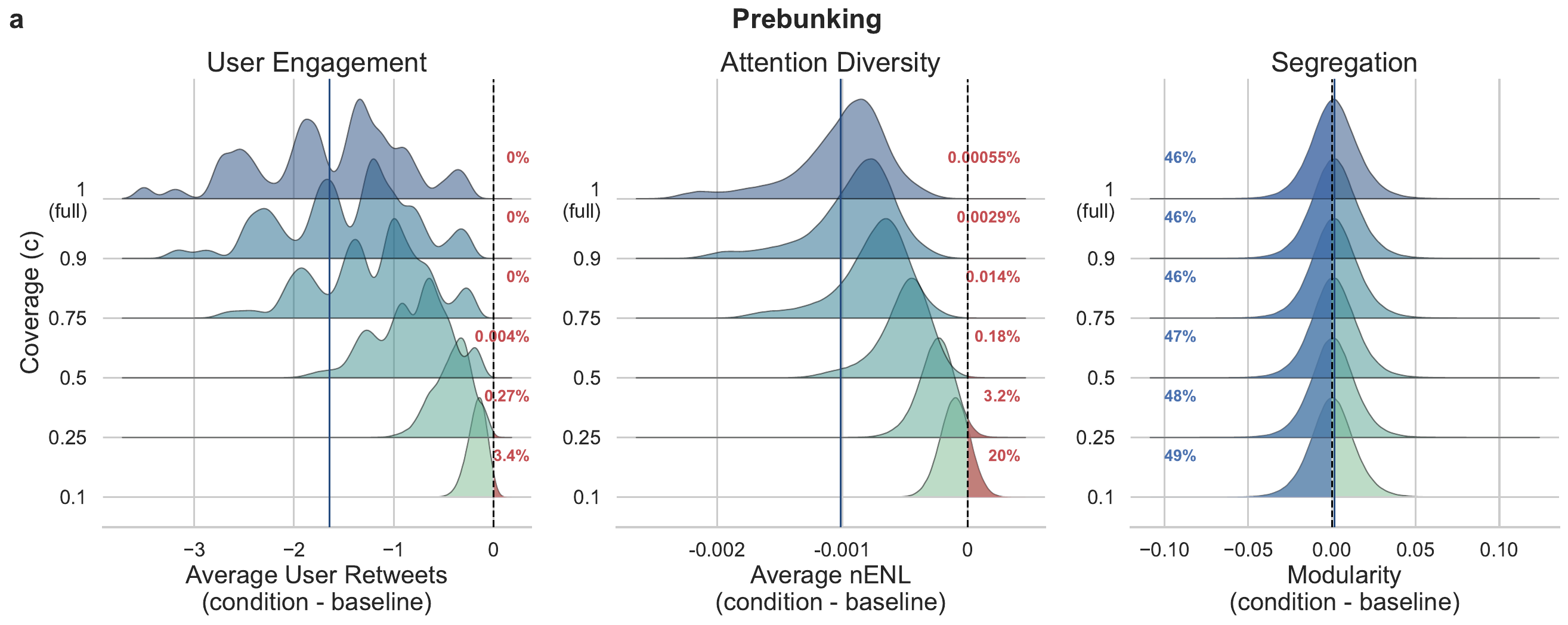}
    \end{subfigure}%
    
    \vspace{0.1cm}
    
    \begin{subfigure}[t]{0.9\textwidth}
        \centering
        \includegraphics[width=\textwidth]{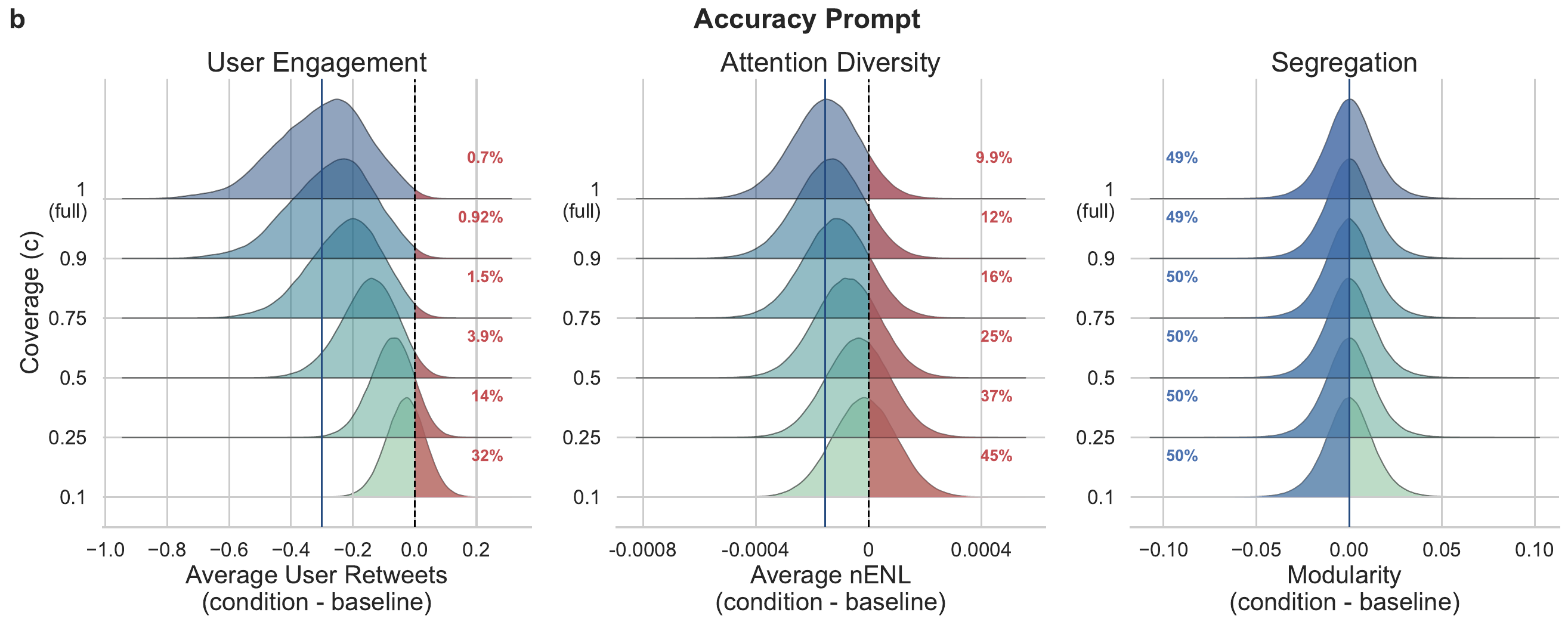}
    \end{subfigure}

     \vspace{0.1cm}

    \begin{subfigure}[t]{0.9\textwidth}
        \centering
        \includegraphics[width=\textwidth]{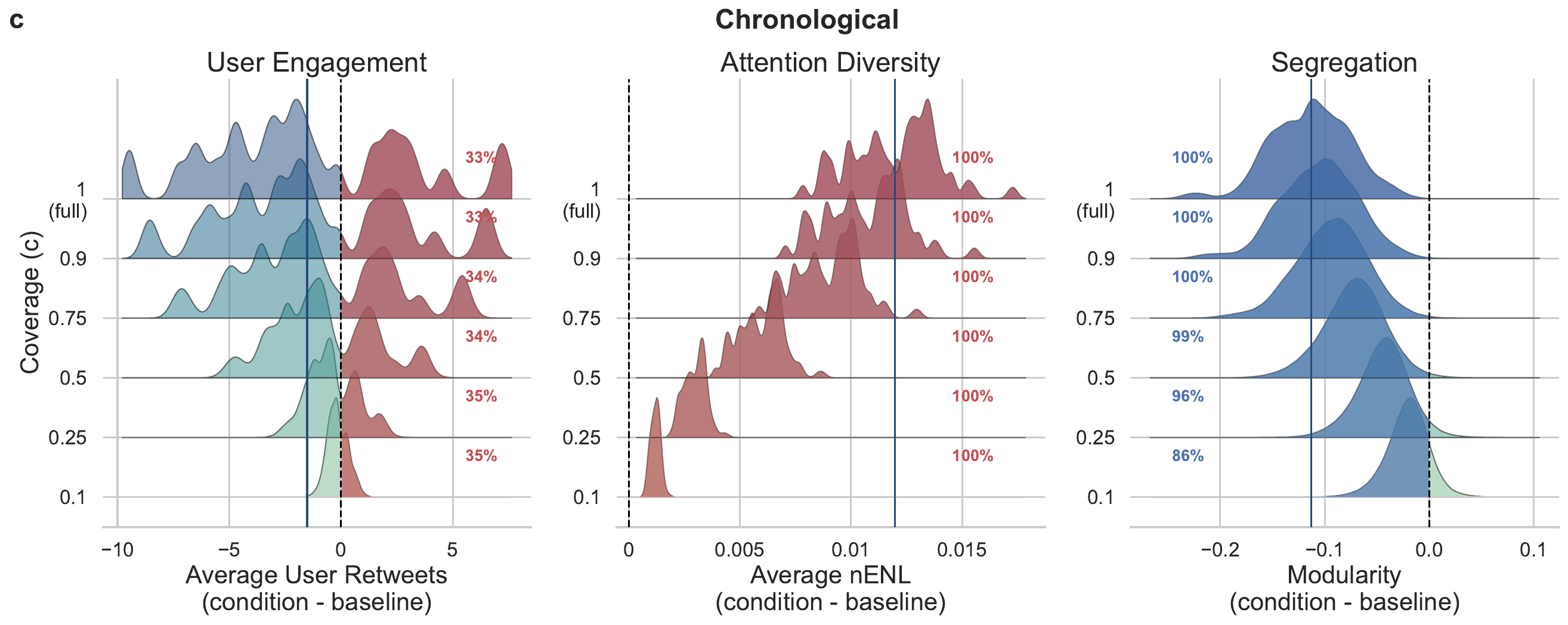}
    \end{subfigure}

    \vspace{0.1cm}

    \begin{subfigure}[t]{0.9\textwidth}
        \centering
        \includegraphics[width=\textwidth]{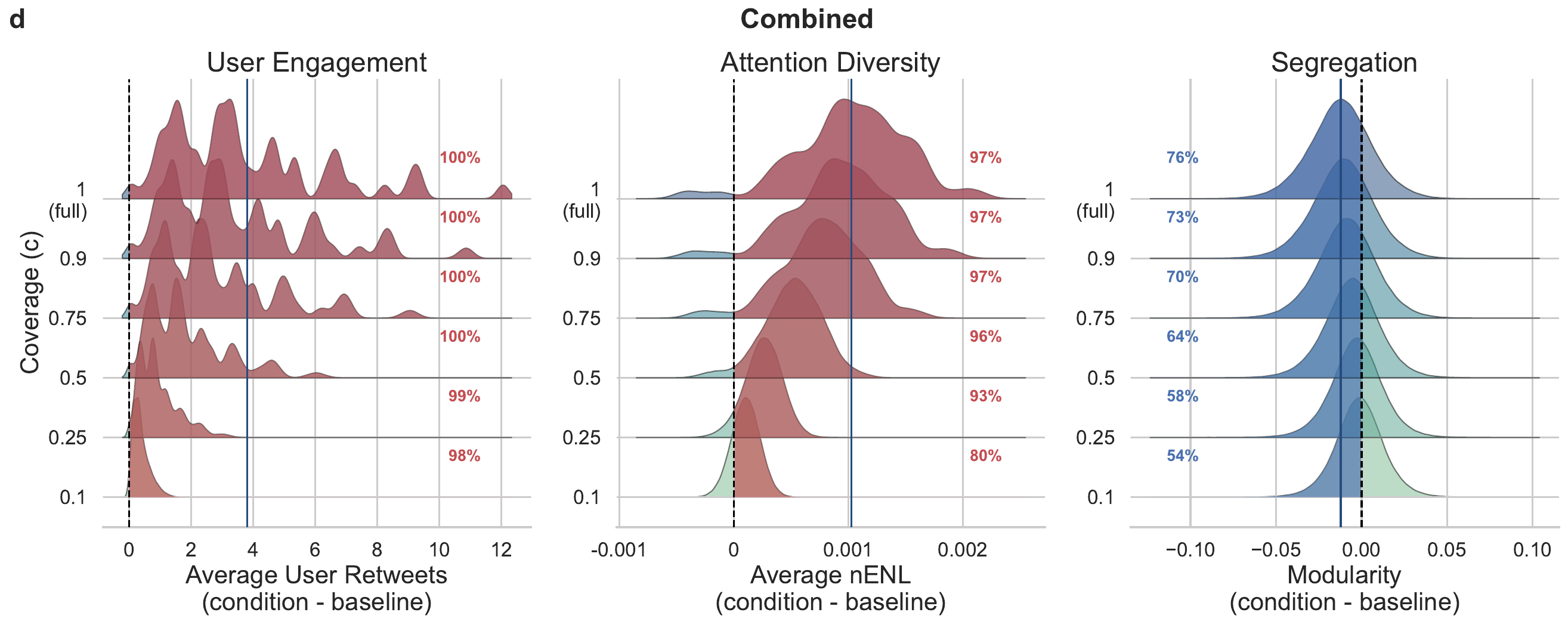}
    \end{subfigure}
    \addtocounter{figure}{-1}
\end{figure}

\begin{figure}[t]
    \caption{Distributions of intervention effects on the system-level metrics as a function of coverage. Each panel shows one intervention: \textbf{a}, prebunking (Cohen's $h = 0.13$); \textbf{b}, accuracy prompts ($k = 0.9$), both applied only to content from non-trustworthy leaders; \textbf{c}, chronological feed; \textbf{d}, combined algorithmic intervention ($\alpha^{\text{INT}} = \alpha^{\text{OPT}}/2$, $\gamma^{\text{INT}} = \gamma^{\text{OPT}}/2$), both applied to all leaders. Columns report user engagement (average user retweets), attention diversity (average nENL) and segregation (modularity). Each ridge is a kernel density estimate of the difference between the metric under the intervention at coverage $c$ and under the baseline (fitted model without intervention, $c = 0$), pooled over the 55 monthly transitions (February to December, 2018--2022) and over the $10{,}000$ Monte Carlo realisations per transition and condition; for $c < 1$, each realisation redraws the set of targeted users. The dashed black line marks zero difference, and the solid blue line marks the mean difference at full coverage ($c = 1$). Shaded areas and percentages give the probability of improvement ($\mathrm{PoI}$): an increase in user engagement and attention diversity (red) or a decrease in segregation (blue). Horizontal scales differ between panels.}
    \label{fig:intervention_assessment}
\end{figure}

\paragraph{Accuracy prompts} Following the same procedure, we assess the intervention examined in the meta-analysis of Pennycook and Rand~\cite{pennycook_accuracy}, in which brief tasks designed to direct users' attention toward content accuracy were administered prior to sharing decisions. The analysis reports that accuracy prompts reduced sharing intentions for misinformation by approximately $10\%$ relative to baseline (thus reducing the average probability of retweet by a factor $k=0.9$), with negligible effect on mainstream news. This effect corresponds to a smaller shift in $\delta$ than prebunking (see Methods), and produces qualitatively similar but smaller outcomes (Fig.~\ref{fig:intervention_assessment}b): a decrease in user engagement ($-0.3$ retweets per user) and attention diversity ($-0.0002$ nENL) in most realisations ($\mathrm{PoI}=0.7\%$ and $9.9\%$, respectively), with negligible effect on segregation ($\mathrm{PoI}=49\%$).

\subsubsection{Algorithmic intervention: the chronological feed}\label{sec:results_chronological_intervention}
Gauthier et al.~\cite{chronological_intervention} conducted a field experiment on X in which active users were randomly assigned to either a platform-curated or a chronological feed for seven weeks. The study finds that switching from the chronological to the platform-curated feed increases engagement and shifts political opinions toward more conservative positions, while the reverse switch, from the platform-curated to the chronological feed, produces no comparable effects; neither switch affects affective polarisation or partisanship.
Analysing feed content, the authors find that the platform-curated feed, which also includes posts from accounts users do not follow, favours conservative content and political activists over traditional news outlets. Its effects persist through account following: users exposed to it began following conservative activist accounts and kept them after switching to the chronological feed, which helps explain the asymmetry.

To simulate the switch from the platform-curated to the chronological feed in our framework, we first define a user as a follower of a leader if the user retweeted that leader in the previous month, and is therefore eligible to be exposed to their content in the following month under the chronological feed scenario. The chronological feed is then modelled by setting the exposure stage parameters to $\alpha \to \infty$, $\beta \to 0$ and $\gamma = 1$, so that each user views content exclusively from leaders they have previously retweeted, with probability proportional to the leader's current activity (see Methods). The inertia parameter $\delta$ remains as fitted to the observed data, since the intervention targets platform curation rather than user behaviour; the viewing parameters $a$ and $b$ are also kept as fitted, assuming that the switch does not change the number of posts users view.

At full coverage, the chronological feed reduces user engagement by $1.5$ retweets per user on average, with only $33\%$ of realisations showing higher engagement than the baseline (Fig.~\ref{fig:intervention_assessment}c). This is in line with the direction reported by Gauthier et al.~\cite{chronological_intervention}, where the effect on user engagement was mild and did not reach statistical significance. The share of realisations with higher engagement stays at $33$--$35\%$ across coverage levels. If the sign were driven by the simulation's randomness (which users are targeted and which retweets are sampled), this share would change with coverage, as for behavioural interventions ($\mathrm{PoI}=3.4\%$ at $c = 0.1$, $\mathrm{PoI}=0\%$ at $c = 1$ in the prebunking analysis). Its stability instead suggests that the sign is set by the month: in about a third of the months, the chronological feed increases engagement.

Attention diversity increases substantially over the baseline, with an average gain of $0.012$ nENL at full coverage, an order of magnitude larger than the changes produced by the behavioural interventions.
The gain occurs in every realisation at every coverage level and grows with coverage.

Segregation also decreases in every realisation at full coverage, with an average reduction in modularity of $0.11$. Already at coverage $c = 0.1$, $86\%$ of realisations show a decrease compared with the baseline, indicating that the community structure of the debate responds even to partial deployment.

These results complement the findings of Gauthier et al.~\cite{chronological_intervention}: whereas they measured affective polarisation and partisanship at the individual level and found no significant effect of switching to the chronological feed, our analysis captures a consistent reduction in the structural segregation of the debate, an effect that becomes evident only when the system is studied as a whole. Since under the chronological feed users can only see leaders they already retweeted, these effects arise entirely from a redistribution of attention within existing user--leader relations: without algorithmic curation, users' engagement, though reduced on average, spreads more evenly across the leaders they already follow instead of concentrating on those favoured by similar users.

\subsubsection{Designing a pro-social algorithmic intervention}\label{sec:results_combined_intervention}
In the previous sections, we considered specific real-world interventions, namely behavioural nudges and the transition to a chronological feed. While useful, both types of intervention exhibited undesirable side effects: all three interventions reduced average user engagement, and behavioural interventions also reduced attention diversity while leaving segregation essentially unchanged.

We now show how the modelling framework can be used to actively design interventions that achieve favourable outcomes across all three properties. By jointly modulating multiple model parameters, we identify a targeted combination that increases attention diversity and reduces segregation while also increasing engagement.

To guide this design, we first conducted a preliminary analysis by varying each model parameter individually (Supplementary Information).
This analysis reveals two complementary effects. Decreasing the selectivity parameter $\alpha$ reduces segregation with negligible effects on attention diversity, but also reduces user engagement, likely because giving more weight to dissimilar users diverts exposure towards leaders the user is less likely to retweet. Conversely, decreasing $\gamma$ has little effect on segregation, but increases user engagement and, more moderately, attention diversity by suppressing the bias toward highly active leaders and refocusing the feed on users' preferred content. These complementary effects motivate a demonstrative combined intervention setting $\alpha^{\text{INT}} = \alpha^{\text{OPT}}/2$ and $\gamma^{\text{INT}} = \gamma^{\text{OPT}}/2$.

At full coverage, this intervention moves all three metrics in the favourable direction: user engagement ($+3.8$ retweets per user), attention diversity ($+0.001$ nENL), and segregation ($-0.01$ modularity), with improvements over the baseline in $100\%$, $97\%$ and $76\%$ of realisations, respectively (Fig.~\ref{fig:intervention_assessment} d). Both the effects and the share of improving realisations grow with coverage.

Compared with the chronological feed, this gain in engagement comes with much smaller improvements in attention diversity and segregation, about one tenth of the average effect on both.

These findings demonstrate that the modelling framework can be used not only to assess known interventions but also to design novel ones, directly targeting the global metrics of interest, with practical implications for platform design. In this case, the engagement gain originates from the exposure re-allocation induced by lowering $\gamma$, which redirects attention toward leaders with whom users hold stronger established relations.

Within the model's assumptions, these results suggest that reducing segregation and diversifying attention need not require algorithmic changes that lower engagement, but may instead be achievable through calibration of existing recommendation parameters. The comparison with the chronological feed indicates, however, that larger improvements in diversity and segregation may come at some cost in engagement, a trade-off that the framework makes it possible to explore.

\section{Discussion}\label{discussion}

In this work we proposed the Exposure--Engagement model, a parametric model that separates content exposure and user engagement into two interpretable stages, and applied it to the Italian public debate on Twitter between 2018 and 2022. Conditioned on the previous month and on the leaders' observed activity, the model describes users' retweet behaviour better than an inertial baseline in 54 of 55 months and tracks three system-level properties we study (user engagement, attention diversity and segregation), well enough to support comparisons between simulated conditions.
On these properties, we quantified the systemic effects of three real-world interventions and one designed intervention by simulating counterfactual scenarios for the Italian Twitter debate.

The behavioural interventions, prebunking and accuracy prompts, act on engagement with content from non-trustworthy sources; at the effect sizes reported in the original experiments~\cite{real_intervention_cohen, pennycook_accuracy}, they modestly reduced user engagement and attention diversity and left segregation essentially unchanged. The chronological feed, by contrast, acts on exposure: at full coverage, it increased attention diversity and reduced segregation in all simulations, at a cost in engagement comparable to that of prebunking.
This engagement cost agrees in direction with field experiments on chronological feeds for US users, on Facebook and Instagram~\cite{guess2023social} and, without reaching statistical significance, on X~\cite{chronological_intervention};
the agreement on experiments with US users and the model fitted to Italian data suggests that the direction of the effect on engagement is not specific to the US context, although it does not constitute a replication.
Our results on attention diversity and segregation have no direct experimental counterpart, because field experiments measure individual outcomes, whereas these are properties of the debate as a whole. Taken together, the results suggest that, for the interventions studied here, changing what users see does more for the collective structure of the debate than changing how they respond to it.
Part of this contrast may reflect differences in effect size and in reach: the behavioural interventions, designed to counter misinformation, have small experimental effect sizes and act only on content from non-trustworthy sources, whereas the exposure-side interventions act on all leaders.

Within the model, an intervention can also improve all three properties at once. Guided by a single-parameter sensitivity analysis (Supplementary Information),
we halved $\alpha$ and $\gamma$. Lowering $\gamma$ redirects exposure from the most active leaders towards those with whom users hold stronger relations, raising engagement and content diversity; lowering $\alpha$ gives relatively more weight to dissimilar users in determining what each user is shown, which reduces segregation but, on its own, also engagement. Combined, the two changes increased engagement in all simulations at full coverage and improved attention diversity and segregation in 97\% and 76\% of them, respectively. These gains in diversity and segregation, however, are about one tenth of those of the chronological feed: the designed intervention removes the engagement cost, but the trade-off persists in the size of the pro-social gains. The factor of 2 is illustrative rather than optimised, and whether real users would respond as the model predicts can only be established through deployment. The result nonetheless identifies a candidate design that platforms could test without an engagement cost, and shows how the framework can screen ranking designs before field experiments.

These results are relevant to current regulatory trends, which increasingly target exposure: the Digital Services Act requires very large online platforms to offer at least one recommender option not based on profiling~\cite{DSA2022}, and the European Commission's guidelines on the protection of minors ask platforms to assess whether such an option should be the default for minors~\cite[Section~6.5.2]{dsa_minors_guidelines2025}. Because such measures rarely reach every user, their system-level effects depend on coverage, which is set by whether an option is opt-in or the default, by the population a rule targets and by uptake and compliance. Even when only 10\% of users received it, the chronological feed reduced segregation in 86\% of simulations. Coverage was random in our simulations, however, whereas uptake and age-based rules do not select users randomly, and the population we study is highly active and most likely adult: these results show how effects can scale with coverage, not what a specific regulation would achieve. The Digital Services Act also requires very large platforms to assess and mitigate systemic risks to civic discourse, including those arising from their recommender systems~\cite{DSA2022}. A model fitted to platform data offers a transparent and reproducible way to compare candidate mitigations before deployment, provided that researchers can access such data, which has become harder as platforms restricted their research interfaces~\cite{davidson2023platform} and enforcement of the Digital Services Act's data-access provisions has been slow~\cite{ec2025xfine}.

Our approach complements existing ways of assessing interventions, which focus on individual-level effects: field experiments estimate how an intervention affects the users who receive it~\cite{guess2023social, guess2023reshares, chronological_intervention}, and recent work suggests that large language models can anticipate the results of survey experiments~\cite{ashokkumar2026large}. The closest modelling study simulated aggregate engagement volumes, without network structure or an exposure mechanism~\cite{bak2022combining}. Here, interventions act on users embedded in the empirical follower--leader network under an explicit model of exposure, so that their effects include the structural reorganisation of the debate. Fitting the model to Italian data also extends this line of research beyond the US, which dominates the empirical literature~\cite{lorenz2023systematic}, to a multi-party context in which a single left--right partisanship score is harder to apply. Measuring segregation as the modularity of the leader-similarity network avoids the need for such labels.

The model has several limitations. It conditions each month on the previous one and on the leaders' observed activity, so it does not represent exogenous shocks that change users' behaviour directly, as its reduced performance in March 2020, at the start of Italy's first COVID-19 lockdown, illustrates; incorporating exogenous signals could address this. Its engagement stage assigns zero retweet probability to follower--leader pairs with no interaction in the previous month, so simulated matrices can lose but never gain relations. This plausibly contributes to the underestimation of user engagement and attention diversity and the overestimation of segregation found in the validation, and implies that the effects of the chronological feed arise solely from reallocating attention within existing relations. The model also holds leaders' activity fixed and does not propagate effects between users or across months, so it captures users' immediate response to an intervention rather than longer-run adaptation, such as changes in what leaders post or whom users follow, which can matter~\cite{chronological_intervention}. Translating experimental effect sizes into parameters further assumes that effects measured in other settings transfer to Twitter, persist over a month and map proportionally onto sharing. Finally, our filtering protocol retains the consistently active core of the debate, whose responses may differ from those of occasional users, and the model assigns the same parameters to all users within a month. Validating the framework on other countries and platforms would test the generality of our findings, and the framework could be extended to further interventions, such as content moderation, targeting based on users' structural role in the network, and policies that remove a subset of users, such as bot removal or the age-based ban introduced in Australia~\cite{australia2024minimumage} and adopted or under consideration in several other countries~\cite{oecd2026agerestrictions}.

Interventions on social media are usually evaluated one at a time, on different platforms and outcomes, and at the level of individuals. By assessing them on the same data and the same system-level properties, the Exposure--Engagement model makes their effects comparable and provides model-based estimates of how they scale with the share of users they reach. For the interventions studied here, the comparison suggests that changing what users are shown does more for the collective structure of the debate than changing how they respond to it. As regulation increasingly targets the design of recommender systems, data-grounded assessments of this kind can help design interventions and anticipate their systemic effects before they are deployed.

\section{Methods}\label{sec:Methods}

\subsection{Dataset description}\label{sec:dataset}

We analysed tweets published on Twitter between 2018 and 2022 by a curated selection of Italian accounts, including politicians, parties and media outlets. The accounts were selected from lists of Italian news outlets and political actors provided by NewsGuard and the Italian Communications Authority (AGCOM), whose monitoring of the national information system spans about $1{,}800$ news sources across media~\cite{agcom2018}, so that the selection spans the mainstream Italian public debate on Twitter.

A subset of 99 out of 583 leaders was annotated as non-trustworthy using the same source classification as a previous study of the Italian news ecosystem~\cite{gravino2022supply}: a list of questionable outlets that AGCOM compiled by merging those of independent fact-checking organisations (bufale.net, butac.it, facta.news and pagellapolitica.it), whose methodologies it verified through their recognition by international fact-checking bodies. As in that study, the annotation is assumed to hold throughout the observation period. These leaders are the targets of the behavioural interventions.

We collected the leaders' tweets that received at least 20 retweets or 10 quotes and gathered all the retweets and quotes they received from other accounts (only retweets enter the model; Section~\ref{sec:Model}). We refer to these accounts as ``users'' or ``followers''; the latter term does not imply that they follow the leaders on the platform, but distinguishes them from the leaders, who are also Twitter users.

For each tweet, quote and retweet, we recorded its identifier, its author and, where applicable, the source tweet it refers to, as well as the creation date of tweets. Because retweets lack native timestamps in our collection, each retweet was assigned the creation date, and hence the timeframe, of the tweet it refers to, since most retweets occur within hours of the original tweet~\cite{kwak2010twitter}.

\begin{table}
    \centering
    \begin{tabular}{llrrrr}
        \toprule
        Dataset & & Followers & Leaders & Tweets & Retweets \\
        \midrule
        Raw, total      &  & 2.713.443 & 583 & 13.996.467 & 29.131.371 \\
        Raw, monthly      & min & 60.267  & 120 & 191.167 & 282.186   \\
                          & max & 370.322 & 163 & 290.171 & 1.144.919 \\
        \addlinespace
        Filtered, monthly & min & 7.393   & 75  & 66.099  & 107.884   \\
                          & max & 9.432   & 90  & 137.427 & 386.840   \\
        \bottomrule
    \end{tabular}
    \caption{The size of the raw and filtered dataset. Since the analysis is conducted monthly, we also report the minimum and maximum size of the monthly sub-datasets.}
    \label{tab:dataset}
\end{table}

We analysed the data at monthly resolution. Because the Exposure--Engagement model maps each month's retweet matrix onto the next, the same users and leaders must be present in consecutive months. We therefore filtered each calendar year separately, retaining leaders who, for each year, published at least ten tweets and received at least one retweet in every month of the year, and users who retweeted at least once in every month of the year. Each year thus forms a separate dataset with fixed sets of users and leaders, in which January serves only to condition the model for February. Filtering by calendar year is a compromise: requiring activity across all five years would retain far fewer users, whereas month-specific populations would break the correspondence between consecutive matrices. The resulting datasets, raw and filtered, are summarised in Table~\ref{tab:dataset}.

\subsection{Probabilistic modelling framework}\label{sec:Model}
We model a social media system in which every account can publish posts and interact with other accounts' posts. Accounts are divided into two categories: $M$ ``leaders'', i.e., influential accounts, and $N$ ``users'' or ``followers'', i.e., the other accounts that interact with leaders' posts. Publication and interaction can happen at any moment in time, but in our modelling scheme aggregates them over $T$ ``timeframes''. Each timeframe is a period long enough for significant relations to emerge in the interaction patterns, which, in the present work, is set to one month: longer periods might lead to spurious interaction patterns as relations change, while shorter periods might not capture enough statistics. We focus on retweet interactions, since they are more abundant than other types (e.g., quotes) and less ambiguous, as a retweet is generally considered an endorsement of the follower toward the leader; the model is nevertheless general enough to accommodate other interaction types without formal changes.

For a timeframe $t$, we introduce the leader activity $A_l(t)$, i.e., the number of posts leader $l$ made in timeframe $t$, collected in the $M$-dimensional vector $\boldsymbol{A}(t)$, and the number of retweets $r_{ul}(t)$ user $u$ made to leader $l$ in the same timeframe, collected in the $N \times M$ matrix $\boldsymbol{r}(t)$. The purpose of the framework is to characterise the users' response to the leaders' activity in the next month, given the leaders' activity in that month and the status of the system in the current one:
\begin{equation*}
    (\boldsymbol{A}(t+1),\boldsymbol{A}(t),\boldsymbol{r}(t)) \rightarrow \boldsymbol{r}(t+1).
\end{equation*}
The Exposure--Engagement model is a parametrised probabilistic model for the random variable $R_{ul}(t)$, of which $r_{ul}(t)$ represents a specific realisation; starred values, such as $r^*_{ul}(t)$, indicate values of observed or known data. The generic leader's activity $A_l(t)$ is not modelled as a random variable: it is always measured from real data and will not be starred. The model is governed by six real-valued parameters, which control distinct mechanisms of the exposure and engagement stages introduced in the Results; realistic parameter values are inferred by minimising the discrepancy metric defined in Section~\ref{sec:Parameter_Optimization}, and interventions are simulated by altering the fitted values.

\subsubsection{The Exposure stage}
We define $\nu_u(t)$ as the number of posts user $u$ sees in timeframe $t$. Given that $u$ will see $\nu_u(t)$ tweets in timeframe $t$, we need to describe how they are distributed across all posts produced in the timeframe $t$ by the $M$ leaders, i.e., $\boldsymbol{A}(t)$. Users may also see and retweet posts from other accounts that are not monitored, but our model focuses only on the considered leaders, assuming that interactions with the rest of the network do not meaningfully affect the system under study.

If there were no preference-based recommendations, the platform would randomly distribute user $u$'s visualisations across the posts produced by the leaders during the timeframe. For the sake of tractability, we assume that this sampling is done with replacement, and we have a multinomial process:
\begin{equation}
    \mathrm{Mult}\left(\nu_u(t),\frac{\boldsymbol{A}(t)}{\sum_{j=1}^M A_{j}(t)}\right)
    \label{eq:chrono-exposure}
\end{equation}
since the probability to visualise tweets from leader $l$ would simply be $\frac{A_l(t)}{\sum_{j=1}^M A_{j}(t)}$. In this random scenario, the probability of viewing a leader's post is proportional to their activity. In real-world scenarios, two factors must be taken into account. First, the impact of a leader's activity on visualisations is not necessarily ``linear'', so we parametrise the influence of posting frequency through the concept of \textit{effective activity}, defined as $A_l(t)^{\gamma}$. We refer to the parameter $\gamma$ as the \textit{leader influence}: it modulates the effect of leaders' posting activity on content exposure, with larger values favouring the visibility of content from the most active leaders.

Second, recommender systems bias each user towards different leaders, an effect we quantify through $\omega_{ul}(t)$. Combining this with the concept of \textit{effective activity}, user $u$ sees a post from leader $l$ at time $t$ with probability $q_{ul}(t) := \frac{A_l(t)^{\gamma} \omega_{ul}(t)}{\sum_{j=1}^M \omega_{uj}(t)A_j(t)^{\gamma}}$. The exposure of users to the leaders' activity is therefore described by
\begin{equation}
    \mathrm{Mult}\left(\nu_u(t),\{q_{ul}(t)\}_{l=1}^M\right).
\end{equation}

\subsubsection{The Engagement stage and the resulting stochastic process}\label{subsec:engage_stochprocess}
We assume that, for each viewed post, user $u$ makes an independent decision based on its author: a post by leader $l$ is engaged through a retweet with probability $p_{ul}$. We call this quantity the ``individual retweet probability''; it represents the relationship between user $u$ and leader $l$, independent of $l$'s activity or $u$'s visualisations.

The process for the individual user $u$, for each timeframe, can thus be conceptualised as $\nu_u$ independent trials, where each trial has $M+1$ possible outcomes: either retweeting a post by one of the $M$ leaders, with probability $\pi_{ul}:= q_{ul} \, p_{ul}$ for each $l \in \{1,\ldots,M\}$, or viewing but not retweeting any post, with probability $\pi_{u0}:= 1-\sum_{l=1}^M \pi_{ul}$. Accordingly, the outcome count $r_{u0} := \nu_u - \sum_{l=1}^M r_{ul}$ denotes the number of viewed posts that were not retweeted. Again, for the sake of tractability, we assume that the sampling is done with replacement, even if in reality users cannot retweet a leader's post more than once. The complete exposure and retweeting process for the individual user therefore corresponds to a multinomial process
\begin{equation}\label{eq:ee_model}
   \mathrm{Mult}(\nu_u(t), (\pi_{ul}(t))_{l=0}^M).
\end{equation}
Finally, we assume that each user's process, conditioned on the previous timeframe, is independent of the other users' behaviour in the same timeframe. This assumption, together with the above definition of $q_{ul}(t+1)$, allows us to write
\begin{equation}\label{eq:ee_model_dist}
\begin{aligned}    
    P&(\boldsymbol{R}(t+1) = \boldsymbol{r} \mid \boldsymbol{R}(t) = \boldsymbol{r}^*(t) )
    = \prod_{u=1}^{N} P(\boldsymbol{R}_u(t+1) = \boldsymbol{r}_u \mid \boldsymbol{R}(t) = \boldsymbol{r}^*(t) ) \\
    &= \prod_{u=1}^{N}\binom{\nu_u(t+1)}{r_{u0}, \ldots, r_{uM}} \, \pi_{u0}(t+1)^{r_{u0}} \prod_{l=1}^M \left( q_{ul}(t+1) p_{ul}(t+1) \right)^{r_{ul}} \\
    &=\prod_{u=1}^{N}\binom{\nu_u(t+1)}{r_{u0}, \ldots, r_{uM}} \, \pi_{u0}(t+1)^{r_{u0}} \prod_{l=1}^M \left( \frac{A^\gamma_l(t+1) \omega_{ul}(t+1)}{\sum_{j=1}^M \omega_{uj}(t+1) A^\gamma_j(t+1)} p_{ul}(t+1) \right)^{r_{ul}},
\end{aligned}
\end{equation}
which describes the stochastic process of the Exposure--Engagement model for all users at timeframe $t+1$ given timeframe $t$.
In this equation, the leaders' activities $\boldsymbol{A}$ are determined from real data, while $\gamma$ is one of the free parameters to be determined by the model fitting optimisation. $\pi_{u0}(t+1)$ is derived once $q_{ul}(t+1)$ and $p_{ul}(t+1)$ for $l=1,\ldots, M $ are calculated. 
$\nu_u$, $p_{ul}$ and $\omega_{ul}$ are not directly observable, therefore we estimate them from the leaders activities $\boldsymbol{A}$ and the retweet matrix of the previous timeframe $\boldsymbol{r}^*(t)$.

\subsubsection{Estimating the number of visualised posts}
The exposure stage is the most elusive, as visualisation leaves no direct traces in our data. Still, we estimate the number of tweets a user $u$ visualises in timeframe $t+1$ as
\begin{equation}
    \nu_u(t+1) =[a\sum\limits_{l=1}^Mr^*_{ul}(t)+b],
    \label{eq:posts_viewed}
\end{equation}
where the round function $[\cdot]$ ensures that the number is an integer.
That is, $u$ is assumed to view $b$ posts by default, plus $a$ additional posts for each retweet they made in the previous time period. The two parameters, $a$ and $b$, are determined by model fitting optimisation and assumed to be strictly positive. 

Eq.~\eqref{eq:posts_viewed} rests on two assumptions: (i) the level of attention is roughly constant in time, as we do not expect users to dramatically change their usage habits from one month to the next; and (ii) the number of tweets seen is a monotonic non-decreasing function of past retweets, for which we chose the linear form as the simplest and most tractable option. We named $a$ the \textit{effective view-to-retweet ratio}, since it determines how the volume of tweets viewed in a given timeframe scales with past retweet activity, while $b$ is the \textit{baseline viewing}, as it sets the default number of posts viewed by all users, independent of their activity level.

\subsubsection{Estimating the algorithmic biases}
The parameter $\omega_{ul}$ represents the algorithmic bias that influences content exposure from leader $l$ to user $u$, which we assume is determined by an underlying recommender system. In standard implementations of content recommendation algorithms, this bias is typically quantified through similarity scores between users and content sources~\cite{Score1, Score2}. We extend this framework by introducing a parametrised form for these bias weights, building on a previous approach~\cite{Bellina}. We assume that these biases can be estimated from the previous month's user engagement, as expressed by the retweet matrix $\boldsymbol{r}^*(t)$. Therefore, in our modelling framework, the functional representation for the bias is defined as
\begin{equation}\label{eq:weights}
\omega_{ul} (t+1) :=  \sum_{v=1}^{N} s_{uv}^{\alpha}(t) \left(\frac{r^*_{vl}(t)}{\sum_{j=1}^{M} r^*_{vj}(t)}\right)^{\beta},
\end{equation}
where:
\begin{itemize}
    \item $s_{uv} (t)$ is the similarity between users $u$ and $v$, defined as
    \begin{equation}
        s_{uv}(t) = \frac{\sum_l r^*_{ul} (t) r^*_{vl} (t)}{\sqrt{\sum_l r^*_{ul} (t)^2} \sqrt{\sum_l r^*_{vl} (t)^2}}
    \end{equation}
    and also calculated on the previous month retweet matrix;
    \item $\alpha$ is a \textit{selectivity} parameter, controlling the strength of the influence of user similarity on exposure;
    \item $\beta$ is a \textit{preference reinforcement} parameter that controls the weight of the peers' past preference shares on exposure.
\end{itemize}
Intuitively, this formulation captures two essential mechanisms of social media recommendation systems. First, through the similarity term, the system identifies users similar to $u$ based on past-month retweeting patterns: when $\alpha$ is large, only users with very high similarity to $u$ substantially influence the recommendations; conversely, when $\alpha$ is small, even dissimilar users become relevant. Second, through the preference term, the algorithm more likely recommends content from leaders who have captured a significant share of the attention of users similar to $u$, with $\beta$ controlling how strongly these proportional preferences are emphasised.

\subsubsection{Estimating the individual retweet probability}
The probability that user $u$ retweets content from leader $l$ once exposed to it (namely the individual retweet probability) quantifies the relationship between user $u$ and leader $l$ independently of their activity levels. This quantity is not directly observable either, so we again estimate it from the previous month's retweet matrix. In our framework, we define it as
\begin{equation}\label{eq:retweet_probability}
p_{ul}(t+1) := \left(\frac{r^*_{ul}(t)}{A_l(t)}\right)^{\delta},
\end{equation}
where $\delta$ is the \textit{inertia} parameter, which controls how past retweeting behaviour influences future retweeting probability. This formulation assumes that the likelihood of retweeting is related to the proportion of a leader's content that the user has previously retweeted: when $\delta$ is close to $1$, the probability closely tracks historical engagement rates, while lower values of $\delta$ reduce the influence of past behaviour on future decisions. Note that Eq.~\eqref{eq:retweet_probability} implies $p_{ul}(t+1) = 0$ whenever $r^*_{ul}(t) = 0$: the model thereby assumes that a consistently active user does not begin engaging with a new leader within a single month, so that simulated matrices can lose, but never gain, user--leader relations (see Discussion).

\subsection{Model fitting and parameter optimisation} \label{sec:Parameter_Optimization}
In the Exposure--Engagement model, the modelled retweet matrix used for fitting corresponds to the expected value of the stochastic process defined in Section~\ref{subsec:engage_stochprocess}. That is, for each user-leader pair $(u,l)$, the number of retweets $\tilde{r}_{ul}$ at time $t+1$ is given by
\begin{equation}
\tilde{r}_{ul}(t+1,\boldsymbol{\theta}) = \mathbb{E}_{\boldsymbol{\theta}}[R_{ul}(t+1) | \boldsymbol{R}(t) =\boldsymbol{r^*}(t)] = \nu_u(t+1) \, q_{ul}(t+1) \, p_{ul}(t+1),
\end{equation}
where $\boldsymbol{\theta} = (\alpha, \beta, \gamma, \delta, a, b)$ denotes the free parameters enabling the calculation of the three factors on the right hand side of the equation.

The model is fit to the observations by minimising the sum of absolute logarithmic ratios
\begin{equation}
\text{MALE}_{t+1}(\boldsymbol{\theta}) = \frac{1}{N\, M} \sum_{u,l} \left| \log\left(\frac{1 + r^*_{ul}(t+1)}{1 + \tilde{r}_{ul}(t+1,\boldsymbol{\theta})}\right) \right|.
\label{eq:Objective_function}
\end{equation}
This metric, which we refer to as the Mean Absolute Logarithmic Error ($\text{MALE}$), emphasises relative rather than absolute errors, treating the same proportional error equally regardless of magnitude. Because retweet counts approximately follow a power-law distribution~\cite{bild2015aggregate}, the logarithmic transform projects the data onto a linear scale and prevents highly active user--leader pairs from disproportionately dominating the estimation procedure. Additionally, using absolute deviations rather than squared differences makes the metric more robust to outliers, which are common in the typically skewed social media engagement patterns, and adding $1$ to both observed and expected values addresses the prevalence of zero values in sparse retweet matrices. In this way, the metric appropriately captures performance across both high-engagement and low-engagement interactions.

The optimal parameters are then calculated as
\begin{equation*}
\boldsymbol{\theta}^{\text{OPT}} \in \arg\min_{\boldsymbol{\theta}} \text{MALE}_{t+1}(\boldsymbol{\theta}),
\label{eq:optimization}
\end{equation*}
leading to
\begin{equation}\label{eq:model_prediction}
    \tilde{r}^{\text{OPT}}_{ul}(t+1)
    = \tilde{r}_{ul}(t+1,\boldsymbol{\theta}^{\text{OPT}})
    = \mathbb{E}_{\boldsymbol{\theta}^{\text{OPT}}}[R_{ul}(t+1) | \boldsymbol{R}(t) =\boldsymbol{r^*}(t)]
\end{equation}
We performed the parameter optimisation within each year of the database, on month $t+1$ conditioning on month $t$, yielding five temporal sequences of eleven parameter estimates, February to December, excluding January for being the first month of the year. Parameter estimates differ for each timeframe, which must be taken into account when interpreting them: parameters $\alpha$, $\beta$, $\delta$, and $a$ act on the system state at time $t$ and influence the expected matrix for time $t+1$; in contrast, $\gamma$ directly modulates leader activities $\boldsymbol{A}$ at time $t+1$; while $b$ represents the baseline viewing level at time $t+1$.

Parameter optimisation was conducted using Differential Evolution~\cite{Differential_evolution}, a population-based, gradient-free metaheuristic implemented in the SciPy library, chosen because the objective landscape may contain multiple local minima and the discontinuities introduced by the rounding in Eq.~\eqref{eq:posts_viewed}. Optimisation terminates upon reaching either convergence tolerance or a maximum number of iterations.

To verify that the model's mechanisms add descriptive value over pure persistence, we also compare it against a na\"ive baseline that assumes no change in behaviour:
\begin{equation}
r^{\text{BASE}}_{ul}(t+1) = r^{\text{*}}_{ul}(t) \quad \forall u,l.
\label{eq:baseline}
\end{equation}
Positive values of $\text{MALE}^{\text{BASE}}_{t+1} - \text{MALE}_{t+1}(\boldsymbol{\theta}^{\text{OPT}})$, where
\begin{equation}
    \text{MALE}^{\text{BASE}}_{t+1}= \frac{1}{N \, M} \sum_{u,l} \left| \log\left(\frac{1 + r^*_{ul}(t+1)}{1 + r^*_{ul}(t)}\right) \right|,
\end{equation}
indicate that the model describes the observed matrix more accurately than the baseline, with larger values corresponding to a greater improvement.

\subsection{System-level metrics} \label{sec:metrics}

After optimising the model parameters as described in the previous section, we can generate simulated retweet matrices and compare them with empirical data. This serves both to evaluate the model's ability to reproduce realistic patterns and to assess the impact of interventions on relevant social metrics. In principle, many different structural or microscopic properties could be measured from the generated $\boldsymbol{r}$ of a counterfactual scenario. For this work we focused on three aspects: the engagement intensity, as the platform's sustainability must be taken into account; the diversity of such engagement; and the segregation of the emerging social network.

\subsubsection{User engagement}
For each user $u$ during month $t$, we define their activity level as
$$E_u(t) = \sum_{l=1}^M r_{ul}(t)$$.
This metric quantifies the total number of retweets made by user $u$ in month $t$, and is averaged across the whole population to yield the system-level user engagement. Monitoring engagement is particularly important when designing interventions, as effective strategies should influence attention distribution without substantially reducing overall platform engagement, which is essential for platform sustainability.

\subsubsection{Attention diversity}
To capture diversity patterns during month $t$, we define the normalised retweet matrix $\boldsymbol{x}(t)$, where each element is defined as
\begin{equation}
x_{ul}(t) = \frac{r_{ul}(t)}{E_u(t)}.
\end{equation}
This normalisation transforms raw retweet counts into proportions that sum to $1$ for each user, representing the fraction of user $u$'s attention directed toward leader $l$. A suitable metric that captures how each user's attention is distributed among leaders in timeframe $t$ is then the normalised Effective Number of Leaders (nENL):
\begin{equation}
\text{nENL}_u(t) = \frac{1}{M \,\sum_{l=1}^M x_{ul}^2(t)}
\end{equation}
The $\text{nENL}_u(t)$ ranges from $\frac{1}{M}$, when a user concentrates all retweets on a single leader, to $1$, when retweets are equally distributed among all leaders: it translates a complex distribution into the equivalent fraction of leaders the user would engage with if they divided their attention equally. This user-level quantity is averaged across the entire population to yield the average nENL measuring the global attention diversity.

\subsubsection{Segregation}\label{sec:modularity}
Beyond individual user metrics, we measure segregation as the extent to which leaders' audiences are segregated into weakly-overlapping communities, which we capture by analysing the structural relationships between leaders through a weighted network $G = (V, E)$, where $V$ represents the set of leaders, thus $|V| = M$, and edge weights are defined as the cosine similarity between leaders' retweet patterns:
$$w_{lm} = \frac{\sum_{u=1}^N  r_{ul} r_{um}}{\sqrt{\sum_{u=1}^N  r_{ul}^2} \sqrt{\sum_{u=1}^N  r_{um}^2}}$$
This cosine similarity formulation quantifies the directional similarity between leaders based on their retweet patterns, providing a scale-invariant measure that focuses on the similarity of retweet distributions, irrespective of the absolute volume of retweets.

With this network representation, we detect communities using the Leiden algorithm~\cite{leiden}, which maximises the network's modularity: this quantity measures the density of weighted connections within communities compared to what would be expected in a random network with the same weighted degree distribution. Higher modularity values indicate stronger community structure, in which leaders within the same community have higher cosine similarity with each other than with leaders in other communities; the resulting maximised modularity defines our segregation. Community detection was performed with the Leiden algorithm~\cite{leiden} as implemented in \texttt{python-igraph} (version 1.0.0; \texttt{Graph.community\_leiden}), optimising weighted modularity and iterating until the partition no longer improved.

\subsection{Implementing interventions}
Interventions are implemented through probabilistic targeting of individual users. Let $\mathcal{U_S} \subseteq \{1, \dots, N\}$ and $\mathcal{L_S} \subseteq \{1, \dots, M\}$ denote the sets of targeted users and leaders, respectively.
For a coverage level $c \in [0,1]$, a fraction $c$ of the users in $\mathcal{U_S}$, drawn uniformly at random without replacement, is selected for intervention, while the remaining users keep the optimal parameters found through minimisation of~\eqref{eq:Objective_function}. The selection is redrawn every time a retweet matrix is simulated (see the later section about Monte Carlo simulation).
If selected, we modify the parameters $\boldsymbol{\theta} = (\alpha, \beta, \gamma, \delta)$ for all pairs $(u, l)$ with $l \in \mathcal{L_S}$, while leaving parameters $(a, b)$ unchanged, as justified in the Results section~\ref{res:Intervention_Assessment}. Mathematically, we represent this intervention as:
\begin{equation}\label{eq:intervention}
  \boldsymbol{\theta}_{ul} =
  \begin{cases}
    \boldsymbol{\theta}^{\,\mathrm{INT}}
      & \text{if } u \text{ is selected and } l \in \mathcal{L_S}, \\[6pt]
    \boldsymbol{\theta}^{\,\mathrm{OPT}}
      & \text{otherwise.}
  \end{cases}
\end{equation}

We apply this framework to three families of interventions.
First, we simulate two empirically calibrated behavioural interventions that modify $\delta$ for interactions with non-trustworthy leaders (Section~\ref{sec:results_behavioural_interventions}). Next, we analyse the transition to a chronological feed by setting $\alpha \to \infty, \beta \to 0, \gamma =1$ (Section~\ref{sec:results_chronological_intervention}). Finally, we propose a combined intervention that simultaneously acts on $\alpha$ and $\gamma$, designed to improve diversity and engagement and reduce segregation (Section~\ref{sec:results_combined_intervention}).

\subsubsection{Behavioural interventions: prebunking and accuracy prompts}\label{sec:Methods_behavioural_interventions}
We denote by $\mathcal{L}_{NT} \subset \{1, \dots, M\}$ the subset of leaders annotated as non-trustworthy (Section~\ref{sec:dataset}), a source-level classification approach already used in the literature on disinformation~\cite{grinberg2019fake, gravino2022supply,1fab}.
Then, to simulate behavioural interventions such as prebunking and accuracy prompts, we act on the user's retweeting probability $p_{ul}$ by modifying the inertia parameter $\delta$ for interactions involving leaders in $\mathcal{L}_{NT}$. In particular, we derive a modified inertia parameter, $\delta^{\text{INT}}$, that mimics the aggregate effect size observed in empirical studies.

We first define the set of active user--leader pairs at time $t$ involving non-trustworthy sources as $\mathcal{NT}(t) = \{(u, l) \mid l \in \mathcal{L}_{NT}, r^*_{ul}(t) > 0\}$. Within our framework, the average retweeting probability for contents from leaders in $\mathcal{L}_{NT}$ in timeframe $t$ is obtained using the fitted inertia parameter $\delta^{\text{OPT}}(t)$:
\begin{equation}
\bar{p}_{NT} (t) = \frac{1}{|\mathcal{NT}(t)|} \sum_{(u,l) \in \mathcal{NT}(t)} \left(\frac{r^*_{ul}(t)}{A_l(t)}\right)^{\delta^\text{{OPT}}(t)}.
\end{equation}
An empirical effect size is translated into a target average probability $\bar{p}'_{NT}(t)$ through the appropriate transformation. For instance, an effect size expressed as Cohen's $h$ (e.g. $h = 0.13$ in~\cite{real_intervention_cohen}) yields
\begin{equation}\label{eq:cohen_h_transform}
  \bar{p}'_{NT}(t) \;=\; \sin^2\!\left(\arcsin\sqrt{\bar{p}_{NT}(t)} - \frac{h}{2}\right),
\end{equation}
while a proportional reduction factor $k \in (0,1)$ (e.g. $k = 0.9$ in~\cite{pennycook_accuracy}) yields
\begin{equation}\label{eq:multiplicative_transform}
  \bar{p}'_{NT}(t) \;=\; k \, \bar{p}_{NT}(t).
\end{equation}
Any other mapping from an empirical effect size to a target probability can be used analogously. Finally, we identify the single intervention parameter $\delta^{\text{INT}}$ by numerically solving the following equation for every timeframe $t$:
\begin{equation}
\frac{1}{|\mathcal{NT}(t)|} \sum_{(u,l) \in \mathcal{NT}(t)} \left(\frac{r^*_{ul}(t)}{A_l(t)}\right)^{\delta^{\text{INT}}(t)} = \bar{p}'_{NT}(t)
\end{equation}
In the simulation, $\delta^{\text{INT}}(t)$ replaces $\delta^{\text{OPT}}(t)$ exclusively for the retweeting probabilities of users selected for intervention when interacting with leaders labelled as non-trustworthy in $\mathcal{L}_{NT}$.

\subsubsection{Chronological feed intervention} \label{sec:Methods_chronological_intervention}
In the chronological feed scenario, the platform proposes to each user a time-ordered display of content from ``followed'' accounts, i.e. accounts to which the user specifically subscribed. To reproduce this mechanism in our framework, we must introduce a notion of ``following'', an action we do not directly observe: we consider user $u$ to ``follow'' leader $l$ if there is at least one engagement interaction in the previous timeframe, a condition expressed by the indicator function $\mathbbm{1}(r^*_{ul}(t) > 0)$.

Taking this into account, we reproduce the chronological feed scenario by assuming exposure is similar to the random case in Eq.~\eqref{eq:chrono-exposure}, but only for ``followed'' leaders. This happens because timeframe aggregation renders the actual order irrelevant, and the absence of leaders' activity modulation and of algorithmic biases makes the exposure probability proportional to the relative abundance of posts. In terms of our framework parameters, this means:
\begin{itemize}
    \item $\gamma = 1$ makes a leader's visibility strictly proportional to posting volume, without any modulation;
    \item $\alpha \to \infty$ suppresses the influence of peer similarity since $s_{uv}^\alpha \to 0$ for $u \neq v$ provided that no two users have identical retweet patterns, thereby restricting the recommendation bias to the user's own history;
    \item $\beta \to 0$ eliminates the reinforcement of past preferences, treating all followed leaders equally, regardless of the intensity of past interaction.
\end{itemize}
Under these conditions, the weighting term $\omega_{ul}$ simplifies, and the viewing probability converges, for the chronological feed intervention, to the analytical form
\begin{equation}
q_{ul}^{\text{INT}}(t+1) = \frac{A_l(t+1) \, \mathbbm{1}(r^*_{ul}(t) > 0)}{\sum_{j=1}^M A_j(t+1) \, \mathbbm{1}(r^*_{uj}(t) > 0)}.
\end{equation}
This formulation ensures that a user's attention is distributed among the leaders they ``follow'' strictly in proportion to each leader's activity $A_l(t+1)$, thereby removing the recommender system's structural bias.

\subsection{Monte Carlo simulation of baseline and counterfactual scenarios}\label{sec:MC_simulation}

Each realisation of the Exposure--Engagement model is an instance of either the baseline scenario, fitted to the observed data, or a counterfactual one.
Each condition is defined by the model parameters (Eq.~\eqref{eq:ee_model}) assigned to each user–leader pair in each monthly transition (Eq.~\eqref{eq:intervention}): the baseline uses the fitted parameters for all pairs, whereas an intervention modifies them for the targeted users and leaders.
For each condition and each transition $t \to t+1$, we drew $K = 10{,}000$ realisations of the retweet matrix, thus approximating the distribution of the model in Eq.~\eqref{eq:ee_model_dist} by Monte Carlo sampling. Each realisation was obtained by drawing every user's row independently from the corresponding multinomial distribution and discarding the no-retweet outcome. All realisations were conditioned on the empirical retweet matrix of the previous month, $\boldsymbol{r}^*(t)$, and on the observed leader activity $\boldsymbol{A}(t+1)$, and were never propagated to the following month.
The three metrics of Section~\ref{sec:metrics} were then computed on every realisation.

Then, for the intervention assessment, we used the Exposure--Engagement model to simulate a counterfactual scenario at full coverage ($c = 1$) and compare it with the baseline condition. To this end, we drew $K$ realisations of the model with the parameters of that intervention and examined the Monte Carlo distribution of the difference between each metric under the intervention and under the baseline (see Section~\ref{res:Intervention_Assessment}, and Section~\ref{sec:statistical_analysis} for how the differences are formed).

To simulate an intervention at coverage $c < 1$, each realisation first drew a set of $[cN]$ users uniformly at random without replacement, where $[\cdot]$ denotes rounding to the nearest integer; the rows of these users are sampled with the intervention probabilities, and those of all other users with the baseline probabilities. The targeted set is redrawn in every realisation, so that the resulting distributions also reflect the variability of user targeting. For each of the four interventions, we simulated $c \in \{0.1, 0.25, 0.5, 0.75, 0.9, 1\}$, where $c = 1$ corresponds to all users receiving the intervention. Each coverage level was compared with the baseline ($c = 0$), which is common to all interventions.

For the sensitivity analysis that guided the design of the combined intervention (Supplementary Information),
we repeated the simulation at full coverage after multiplying one of the parameters $\alpha$, $\beta$, $\gamma$ or $\delta$ by $1/5$, $1/2$, $2$ or $5$, with all other parameters at their fitted values.

\subsection{Statistical analysis}\label{sec:statistical_analysis}
\paragraph{Comparison with empirical data.} For each of the $55$ monthly transitions (February to December, 2018--2022), each metric of the simulated baseline was summarised by the mean of its $K = 10{,}000$ Monte Carlo values and by its $95\%$ probability interval, delimited by the $2.5$th and $97.5$th percentiles (Fig.~\ref{fig:user_metrics_correlations}). Agreement with the empirical values was quantified by Pearson's correlation coefficient, with two-sided $p$-values, and by an ordinary least-squares fit of the simulated means on the empirical values, whose slope and intercept are reported with $95\%$ confidence intervals (Table~\ref{tab:correlation_metrics_updated}). Because consecutive months are serially dependent, p-values and confidence intervals were adjusted using the effective sample size $n_{eff} = n ( 1 - \rho_x \rho_y ) / (1 + \rho_x \rho_y )$~\cite{bretherton1999effective}, where $n = 55$ and $\rho_x$ and $\rho_y$ are the lag-1 autocorrelations of the empirical and simulated series. These were estimated from the $50$ pairs of consecutive months within the same year, so that no pair spans a year boundary, where the filtered population changes and January is missing. Since the pairs are pooled across years, the estimates also absorb differences in level between years, which makes the adjustment conservative. The p-values were computed with $n_{eff} - 2$ degrees of freedom; the standard errors of slope and intercept were inflated by $\sqrt{(n - 2)/(n_{eff} - 2)}$, and their confidence intervals were taken from the $t$ distribution with $n_{eff} - 2$ degrees of freedom.
Lag-1 autocorrelations computed separately within each year are reported in Supplementary Information.

\paragraph{Intervention effects.} For each intervention, coverage level $c$, system-level metric $G$ and transition to month $t$, we computed the $K=10{,}000$ differences
\begin{equation}
\Delta G_{t,k}(c) = G\big(\boldsymbol{r}^{(c)}_{t,k}\big) - G\big(\boldsymbol{r}^{\mathrm{BASE}}_{t,k}\big), \qquad k = 1, \dots, K,
\label{eq:MC_differences}
\end{equation}
between the $k$-th realisation under the intervention and the $k$-th, independently generated, realisation of the baseline. Pooling over transitions yields $55K = 550{,}000$ differences for each intervention, coverage level and metric, combining within-month stochastic variability with heterogeneity across months. We summarise them by their mean and standard deviation, and by the Probability of Improvement ($\mathrm{PoI}$), the share of differences in the favourable direction
\begin{equation}
\mathrm{PoI}_G(c) = \frac{1}{55K}\sum_{t}\sum_{k=1}^{K}\mathbbm{1}\big(s \,\Delta G_{t,k}(c) > 0\big),
\label{eq:favourable_share}
\end{equation}
where $s = +1$ when $G$ is the average user engagement and the average nENL, and $s = -1$ when $G$ is the modularity. $\mathrm{PoI}_G(c)$ estimates the probability that a realisation of the intervention improves on the baseline for the same month, averaged over months.
The densities in Fig.~\ref{fig:intervention_assessment} are Gaussian kernel density estimates with bandwidth selected by Scott's rule, rescaled to a common peak height for display.

These quantities are conditional on the fitted parameters, the observed monthly histories and the observed leader activity, and do not include uncertainty from parameter estimation, structural model error or exogenous shocks. We do not report hypothesis tests for intervention effects: because the number of realisations can be increased arbitrarily, any non-zero expected difference would eventually reach statistical significance, so $p$-values would reflect the simulation effort rather than the size of the effect. We therefore interpret $\mathrm{PoI}_G(c)$ as a measure of the consistency of an effect's direction within the model.

\paragraph{Software and reproducibility.} The code to reproduce all results is publicly available at \url{https://github.com/SonyCSLParis/eenterventions.git}.

\bmhead{Acknowledgements}
This work has been supported by the Horizon Europe VALAWAI project (grant agreement number 101070930).

\bibliography{sn-bibliography}

@article{Echo_chamber,
author = {Cinelli, Matteo and Morales, Gianmarco and Galeazzi, Alessandro and Quattrociocchi, Walter and Starnini, Michele},
year = {2021},
month = {03},
pages = {e2023301118},
title = {The echo chamber effect on social media},
volume = {118},
journal = {Proceedings of the National Academy of Sciences},
doi = {10.1073/pnas.2023301118}
}

@article{Loreto_opiniondynamics,
author = {Sîrbu, Alina and Loreto, Vittorio and Servedio, Vito and Tria, Francesca},
year = {2013},
month = {02},
pages = {},
title = {Cohesion, consensus and extreme information in opinion dynamics},
volume = {16},
journal = {Advances in Complex Systems},
doi = {10.1142/S0219525913500355}
}

@article{Disinformation,
author = {Gravino, Pietro and Prevedello, Giulio and Brugnoli, Emanuele},
year = {2024},
month = {07},
pages = {},
title = {Online news ecosystem dynamics: supply, demand, diffusion, and the role of disinformation},
volume = {9},
journal = {Applied Network Science},
doi = {10.1007/s41109-024-00643-1}
}

@article{Toxicity,
  title={Crowdsourced audit of Twitter’s recommender systems},
  author={Bouchaud, Paul and Chavalarias, David and Panahi, Maziyar},
  journal={Scientific Reports},
  volume={13},
  number={1},
  pages={16815},
  year={2023},
  publisher={Nature Publishing Group UK London}
}

@article{Polarization, title={Political Polarization on Twitter}, volume={5}, url={https://ojs.aaai.org/index.php/ICWSM/article/view/14126}, DOI={10.1609/icwsm.v5i1.14126}, abstractNote={ &lt;p&gt; In this study we investigate how social media shape the networked public sphere and facilitate communication between communities with different political orientations. We examine two networks of political communication on Twitter, comprised of more than 250,000 tweets from the six weeks leading up to the 2010 U.S. congressional midterm elections. Using a combination of network clustering algorithms and manually-annotated data we demonstrate that the network of political retweets exhibits a highly segregated partisan structure, with extremely limited connectivity between left- and right-leaning users. Surprisingly this is not the case for the user-to-user mention network, which is dominated by a single politically heterogeneous cluster of users in which ideologically-opposed individuals interact at a much higher rate compared to the network of retweets. To explain the distinct topologies of the retweet and mention networks we conjecture that politically motivated individuals provoke interaction by injecting partisan content into information streams whose primary audience consists of ideologically-opposed users. We conclude with statistical evidence in support of this hypothesis. &lt;/p&gt; }, number={1}, journal={Proceedings of the International AAAI Conference on Web and Social Media}, author={Conover, Michael and Ratkiewicz, Jacob and Francisco, Matthew and Goncalves, Bruno and Menczer, Filippo and Flammini, Alessandro}, year={2021}, month={Aug.}, pages={89-96} }

@article{Giordano,
  title={AI agents can coordinate beyond human scale},
  author={De Marzo, Giordano and Castellano, Claudio and Garcia, David},
  journal={arXiv preprint arXiv:2409.02822},
  year={2024}
}

@article{Bellina,
  title={Effect of collaborative-filtering-based recommendation algorithms on opinion polarization},
  author={Bellina, Alessandro and Castellano, Claudio and Pineau, Paul and Iannelli, Giulio and De Marzo, Giordano},
  journal={Physical Review E},
  volume={108},
  number={5},
  pages={054304},
  year={2023},
  publisher={APS}
}

@article{Score1,
  title={Grouplens: An open architecture for collaborative filtering of netnews},
  author={Resnick, Paul and Iacovou, Neophytos and Suchak, Mitesh and Bergstrom, Peter and Riedl, John},
  journal={Proceedings of the 1994 ACM conference on Computer supported cooperative work},
  pages={175--186},
  year={1994}
}

@article{Score2,
title = {Recommender systems},
journal = {Physics Reports},
volume = {519},
number = {1},
pages = {1-49},
year = {2012},
note = {Recommender Systems},
issn = {0370-1573},
doi = {https://doi.org/10.1016/j.physrep.2012.02.006},
url = {https://www.sciencedirect.com/science/article/pii/S0370157312000828},
author = {Linyuan Lü and Matúš Medo and Chi Ho Yeung and Yi-Cheng Zhang and Zi-Ke Zhang and Tao Zhou}
}

@article{1fab,
author={Emanuele Brugnoli and Pietro Gravino and Donald Ruggiero Lo Sardo and Vittorio Loreto and Giulio Prevedello},
title={Fine-Grained Clustering of Social Media: How Moral Triggers Drive Preferences and Consensus},
journal={Proceedings of the 16th International Conference on Agents and Artificial Intelligence - Volume 3: AWAI},
year={2024},
pages={1405-1412},
publisher={SciTePress},
organization={INSTICC},
doi={10.5220/0012595000003636},
isbn={978-989-758-680-4},
}

@InProceedings{moral_values,
author="Brugnoli, Emanuele
and Gravino, Pietro
and Prevedello, Giulio",
editor="Osman, Nardine
and Steels, Luc",
title="Moral Values in Social Media for Disinformation and Hate Speech Analysis",
booktitle="Value Engineering in Artificial Intelligence",
year="2024",
publisher="Springer Nature Switzerland",
address="Cham",
pages="67--82",
isbn="978-3-031-58202-8"
}

@article{Differential_evolution,
author = {Storn, Rainer and Price, Kenneth},
year = {1997},
month = {01},
pages = {341-359},
title = {Differential Evolution - A Simple and Efficient Heuristic for Global Optimization over Continuous Spaces},
volume = {11},
journal = {Journal of Global Optimization},
doi = {10.1023/A:1008202821328}
}

@article{leiden,
  title={From Louvain to Leiden: guaranteeing well-connected communities},
  author={Traag, Vincent A and Waltman, Ludo and Van Eck, Nees Jan},
  journal={Scientific reports},
  volume={9},
  number={1},
  pages={1--12},
  year={2019},
  publisher={Nature Publishing Group}
}

@article{fake_news_intervention,
  author = {Pennycook, Gordon and Epstein, Ziv and Mosleh, Mohsen and Arechar, Antonio A. and Eckles, Dean and Rand, David G.},
  title = {Shifting attention to accuracy can reduce misinformation online},
  journal = {Nature},
  volume = {592},
  number = {7855},
  pages = {590--595},
  year = {2021},
  doi = {10.1038/s41586-021-03344-2},
  url = {https://doi.org/10.1038/s41586-021-03344-2},
  issn = {1476-4687}
}

@article{real_intervention_cohen,
  title={Psychological inoculation improves resilience against misinformation on social media},
  author={Roozenbeek, Jon and Van Der Linden, Sander and Goldberg, Beth and Rathje, Steve and Lewandowsky, Stephan},
  journal={Science advances},
  volume={8},
  number={34},
  pages={eabo6254},
  year={2022},
  publisher={American Association for the Advancement of Science}
}

@article{pennycook_accuracy,
  title={Accuracy prompts are a replicable and generalizable approach for reducing the spread of misinformation},
  author={Pennycook, Gordon and Rand, David G},
  journal={Nature communications},
  volume={13},
  number={1},
  pages={2333},
  year={2022},
  publisher={Nature Publishing Group UK London}
}

@article{chronological_intervention,
  title={The political effects of X’s feed algorithm},
  author={Gauthier, Germain and Hodler, Roland and Widmer, Philine and Zhuravskaya, Ekaterina},
  journal={Nature},
  pages={1--8},
  year={2026},
  publisher={Nature Publishing Group UK London}
}

@article{lorenz2023systematic,
  title={A systematic review of worldwide causal and correlational evidence on digital media and democracy},
  author={Lorenz-Spreen, Philipp and Oswald, Lisa and Lewandowsky, Stephan and Hertwig, Ralph},
  journal={Nature human behaviour},
  volume={7},
  number={1},
  pages={74--101},
  year={2023},
  publisher={Nature Publishing Group UK London}
}

@article{allcott2020welfare,
  title={The welfare effects of social media},
  author={Allcott, Hunt and Braghieri, Luca and Eichmeyer, Sarah and Gentzkow, Matthew},
  journal={American economic review},
  volume={110},
  number={3},
  pages={629--676},
  year={2020},
  publisher={American Economic Association 2014 Broadway, Suite 305, Nashville, TN 37203}
}

@article{argyle2023out,
  title={Out of one, many: Using language models to simulate human samples},
  author={Argyle, Lisa P and Busby, Ethan C and Fulda, Nancy and Gubler, Joshua R and Rytting, Christopher and Wingate, David},
  journal={Political Analysis},
  volume={31},
  number={3},
  pages={337--351},
  year={2023},
  publisher={Cambridge University Press}
}

@article{park2023generative,
  title={Generative agents: Interactive simulacra of human behavior},
  author={Park, Joon Sung and O'Brien, Joseph and Cai, Carrie Jun and Morris, Meredith Ringel and Liang, Percy and Bernstein, Michael S},
  journal={Proceedings of the 36th annual acm symposium on user interface software and technology},
  pages={1--22},
  doi = {10.1145/3586183.3606763},
  year={2023}
}

@article{bakshy2012role,
  title={The role of social networks in information diffusion},
  author={Bakshy, Eytan and Rosenn, Itamar and Marlow, Cameron and Adamic, Lada},
  journal={Proceedings of the 21st international conference on World Wide Web},
  pages={519--528},
  year={2012}
}

@article{kwak2010twitter,
  title={What is Twitter, a social network or a news media?},
  author={Kwak, Haewoon and Lee, Changhyun and Park, Hosung and Moon, Sue},
  journal={Proceedings of the 19th international conference on World wide web},
  pages={591--600},
  year={2010}
}

@article{rainer2002opinion,
author = {Hegselmann, Rainer and Krause, Ulrich},
year = {2002},
month = {07},
pages = {},
title = {Opinion Dynamics and Bounded Confidence Models, Analysis and Simulation},
volume = {5},
journal = {Journal of Artificial Societies and Social Simulation}
}

@article{degroot1974reaching,
  title={Reaching a consensus},
  author={DeGroot, Morris H},
  journal={Journal of the American Statistical association},
  volume={69},
  number={345},
  pages={118--121},
  year={1974},
  publisher={Taylor \& Francis}
}

@article{carletti2006make,
  title={How to make an efficient propaganda},
  author={Carletti, Timoteo and Fanelli, Duccio and Grolli, Stefano and Guarino, Alessio},
  journal={EPL (Europhysics Letters)},
  volume={74},
  number={2},
  pages={222--228},
  year={2006}
}

@article{crokidakis2012effects,
  title={Effects of mass media on opinion spreading in the Sznajd sociophysics model},
  author={Crokidakis, Nuno},
  journal={Physica A: Statistical Mechanics and its Applications},
  volume={391},
  number={4},
  pages={1729--1734},
  year={2012},
  publisher={Elsevier}
}

@article{vaz2010mass,
  title={Mass media and repulsive interactions in continuous-opinion dynamics},
  author={Vaz Martins, Teresa and Pineda, Miguel and Toral, Raul},
  journal={EPL (Europhysics Letters)},
  volume={91},
  number={4},
  pages={48003},
  year={2010}
}

@article{deffuant2000mixing,
  title={Mixing beliefs among interacting agents},
  author={Deffuant, Guillaume and Neau, David and Amblard, Frederic and Weisbuch, G{\'e}rard},
  journal={Advances in complex systems},
  volume={3},
  number={01n04},
  pages={87--98},
  year={2000},
  publisher={World Scientific}
}

@article{galam2008sociophysics,
  title={Sociophysics: A review of Galam models},
  author={Galam, Serge},
  journal={International Journal of Modern Physics C},
  volume={19},
  number={03},
  pages={409--440},
  year={2008},
  publisher={World Scientific}
}

@article{sznajd2000opinion,
  title={Opinion evolution in closed community},
  author={Sznajd-Weron, Katarzyna and Sznajd, Jozef},
  journal={International Journal of Modern Physics C},
  volume={11},
  number={06},
  pages={1157--1165},
  year={2000},
  publisher={World Scientific}
}

@article{axelrod1997dissemination,
  title={The dissemination of culture: A model with local convergence and global polarization},
  author={Axelrod, Robert},
  journal={Journal of conflict resolution},
  volume={41},
  number={2},
  pages={203--226},
  year={1997},
  publisher={Sage Periodicals Press 2455 Teller Road, Thousand Oaks, CA 91320}
}

@article{lazer2018science,
  title={The science of fake news},
  author={Lazer, David MJ and Baum, Matthew A and Benkler, Yochai and Berinsky, Adam J and Greenhill, Kelly M and Menczer, Filippo and Metzger, Miriam J and Nyhan, Brendan and Pennycook, Gordon and Rothschild, David and others},
  journal={Science},
  volume={359},
  number={6380},
  pages={1094--1096},
  year={2018},
  publisher={American Association for the Advancement of Science}
}

@article{gravino2022supply,
  title={The supply and demand of news during COVID-19 and assessment of questionable sources production},
  author={Gravino, Pietro and Prevedello, Giulio and Galletti, Martina and Loreto, Vittorio},
  journal={Nature human behaviour},
  volume={6},
  number={8},
  pages={1069--1078},
  year={2022},
  publisher={Nature Publishing Group UK London}
}

@misc{DSA2022,
  title = {Regulation ({EU}) 2022/2065 of the {European Parliament} and of the {Council} of 19 {October} 2022 on a Single Market For Digital Services and amending {Directive} 2000/31/{EC} ({Digital Services Act})},
  year = {2022},
  note = {Official Journal of the European Union, L 277, 1--102},
}

@article{grinberg2019fake,
  title={Fake news on Twitter during the 2016 US presidential election},
  author={Grinberg, Nir and Joseph, Kenneth and Friedland, Lisa and Swire-Thompson, Briony and Lazer, David},
  journal={Science},
  volume={363},
  number={6425},
  pages={374--378},
  year={2019},
  publisher={American Association for the Advancement of Science}
}

@misc{agcom2018,
  author       = {{AGCOM, Department of Economics and Statistics}},
  title        = {News vs. fake nel sistema dell'informazione [{News vs. fake in the information system}]},
  howpublished = {Autorit{\`a} per le Garanzie nelle Comunicazioni, Rome},
  year         = {2018},
  month        = nov,
  url          = {https://www.agcom.it/sites/default/files/migration/attachment/Allegato%2025-1-2019_1.pdf},
  note         = {English version. Accessed 20 October 2026}
}

@article{rathje2021out,
  title={Out-group animosity drives engagement on social media},
  author={Rathje, Steve and Van Bavel, Jay J and Van Der Linden, Sander},
  journal={Proceedings of the national academy of sciences},
  volume={118},
  number={26},
  pages={e2024292118},
  year={2021},
  publisher={National Academy of Sciences}
}

@article{castellano2009statistical,
  title={Statistical physics of social dynamics},
  author={Castellano, Claudio and Fortunato, Santo and Loreto, Vittorio},
  journal={Reviews of modern physics},
  volume={81},
  number={2},
  pages={591--646},
  year={2009},
  publisher={APS}
}

@article{centola2010spread,
  title={The spread of behavior in an online social network experiment},
  author={Centola, Damon},
  journal={science},
  volume={329},
  number={5996},
  pages={1194--1197},
  year={2010},
  publisher={American Association for the Advancement of Science}
}

@article{vosoughi2018spread,
  title={The spread of true and false news online},
  author={Vosoughi, Soroush and Roy, Deb and Aral, Sinan},
  journal={science},
  volume={359},
  number={6380},
  pages={1146--1151},
  year={2018},
  publisher={American Association for the Advancement of Science}
}

@article{bail2018exposure,
  title={Exposure to opposing views on social media can increase political polarization},
  author={Bail, Christopher A and Argyle, Lisa P and Brown, Taylor W and Bumpus, John P and Chen, Haohan and Hunzaker, MB Fallin and Lee, Jaemin and Mann, Marcus and Merhout, Friedolin and Volfovsky, Alexander},
  journal={Proceedings of the National Academy of Sciences},
  volume={115},
  number={37},
  pages={9216--9221},
  year={2018},
  publisher={National Academy of Sciences}
}

@article{brady2017emotion,
  title={Emotion shapes the diffusion of moralized content in social networks},
  author={Brady, William J and Wills, Julian A and Jost, John T and Tucker, Joshua A and Van Bavel, Jay J},
  journal={Proceedings of the National Academy of Sciences},
  volume={114},
  number={28},
  pages={7313--7318},
  year={2017},
  publisher={National Academy of Sciences}
}

@article{pennycook2021psychology,
  title={The psychology of fake news},
  author={Pennycook, Gordon and Rand, David G},
  journal={Trends in cognitive sciences},
  volume={25},
  number={5},
  pages={388--402},
  year={2021},
  publisher={Elsevier}
}

@book{wardle2017information,
  title={Information disorder: Toward an interdisciplinary framework for research and policymaking},
  author={Wardle, Claire and Derakhshan, Hossein},
  volume={27},
  year={2017},
  publisher={Council of Europe Strasbourg}
}

@article{lazer2020computational,
  title={Computational social science: Obstacles and opportunities},
  author={Lazer, David MJ and Pentland, Alex and Watts, Duncan J and Aral, Sinan and Athey, Susan and Contractor, Noshir and Freelon, Deen and Gonzalez-Bailon, Sandra and King, Gary and Margetts, Helen and others},
  journal={Science},
  volume={369},
  number={6507},
  pages={1060--1062},
  year={2020},
  publisher={American Association for the Advancement of Science}
}

@book{salganik2019bit,
  title={Bit by bit: Social research in the digital age},
  author={Salganik, Matthew J},
  year={2019},
  publisher={Princeton University Press}
}

@article{kubin2021role,
  title={The role of (social) media in political polarization: a systematic review},
  author={Kubin, Emily and Von Sikorski, Christian},
  journal={Annals of the International Communication Association},
  volume={45},
  number={3},
  pages={188--206},
  year={2021},
  publisher={Oxford University Press}
}

@article{marino2024mis,
  title={(Mis-) Perceptions, information, and political polarization: A survey and a systematic literature review},
  author={Marino, Maria and Iacono, Roberto and Mollerstrom, Johanna},
  journal={European Journal of Political Economy},
  volume={85},
  pages={102578},
  year={2024},
  publisher={Elsevier}
}

@article{hartmann2025systematic,
  title={A systematic review of echo chamber research: comparative analysis of conceptualizations, operationalizations, and varying outcomes},
  author={Hartmann, David and Wang, Sonja Mei and Pohlmann, Lena and Berendt, Bettina},
  journal={Journal of Computational Social Science},
  volume={8},
  number={2},
  pages={52},
  year={2025},
  publisher={Springer}
}

@article{windisch2022online,
  title={Online interventions for reducing hate speech and cyberhate: A systematic review},
  author={Windisch, Steven and Wiedlitzka, Susann and Olaghere, Ajima and Jenaway, Elizabeth},
  journal={Campbell systematic reviews},
  volume={18},
  number={2},
  pages={e1243},
  year={2022},
  publisher={Wiley Online Library}
}

@article{voelkel2024megastudy,
  title={Megastudy testing 25 treatments to reduce antidemocratic attitudes and partisan animosity},
  author={Voelkel, Jan G and Stagnaro, Michael N and Chu, James Y and Pink, Sophia L and Mernyk, Joseph S and Redekopp, Chrystal and Ghezae, Isaias and Cashman, Matthew and Adjodah, Dhaval and Allen, Levi G and others},
  journal={Science},
  volume={386},
  number={6719},
  pages={eadh4764},
  year={2024},
  publisher={American Association for the Advancement of Science}
}

@article{holliday2025depolarization,
  title={Why depolarization is hard: Evaluating attempts to decrease partisan animosity in America},
  author={Holliday, Derek E and Lelkes, Yphtach and Westwood, Sean J},
  journal={Proceedings of the National Academy of Sciences},
  volume={122},
  number={39},
  pages={e2508827122},
  year={2025},
  publisher={National Academy of Sciences}
}

@article{kozyreva2024toolbox,
  title   = {Toolbox of individual-level interventions against online misinformation},
  author  = {Kozyreva, Anastasia and Lorenz-Spreen, Philipp and Herzog, Stefan M. and Ecker, Ullrich K. H. and Lewandowsky, Stephan and Hertwig, Ralph and Ali, Ayesha and Bak-Coleman, Joe and Barzilai, Sarit and Basol, Melisa and Berinsky, Adam J. and Betsch, Cornelia and Cook, John and Fazio, Lisa K. and Geers, Michael and Guess, Andrew M. and Huang, Haifeng and Larreguy, Horacio and Maertens, Rakoen and Panizza, Folco and Pennycook, Gordon and Rand, David G. and Rathje, Steve and Reifler, Jason and Schmid, Philipp and Smith, Mark and Swire-Thompson, Briony and Szewach, Paula and van der Linden, Sander and Wineburg, Sam},
  journal = {Nature Human Behaviour},
  volume  = {8},
  number  = {6},
  pages   = {1044--1052},
  year    = {2024},
  doi     = {10.1038/s41562-024-01881-0}
}

@article{bakshy2015exposure,
  title={Exposure to ideologically diverse news and opinion on Facebook},
  author={Bakshy, Eytan and Messing, Solomon and Adamic, Lada A},
  journal={Science},
  volume={348},
  number={6239},
  pages={1130--1132},
  year={2015},
  publisher={American Association for the Advancement of Science}
}

@article{bak2022combining,
  title={Combining interventions to reduce the spread of viral misinformation},
  author={Bak-Coleman, Joseph B and Kennedy, Ian and Wack, Morgan and Beers, Andrew and Schafer, Joseph S and Spiro, Emma S and Starbird, Kate and West, Jevin D},
  journal={Nature Human Behaviour},
  volume={6},
  number={10},
  pages={1372--1380},
  year={2022},
  publisher={Nature Publishing Group UK London}
}

@article{del2016spreading,
  title={The spreading of misinformation online},
  author={Del Vicario, Michela and Bessi, Alessandro and Zollo, Fabiana and Petroni, Fabio and Scala, Antonio and Caldarelli, Guido and Stanley, H Eugene and Quattrociocchi, Walter},
  journal={Proceedings of the national academy of Sciences},
  volume={113},
  number={3},
  pages={554--559},
  year={2016},
  publisher={National Academy of Sciences}
}

@article{guess2023reshares,
  title={Reshares on social media amplify political news but do not detectably affect beliefs or opinions},
  author={Guess, Andrew M and Malhotra, Neil and Pan, Jennifer and Barber{\'a}, Pablo and Allcott, Hunt and Brown, Taylor and Crespo-Tenorio, Adriana and Dimmery, Drew and Freelon, Deen and Gentzkow, Matthew and others},
  journal={Science},
  volume={381},
  number={6656},
  pages={404--408},
  year={2023},
  publisher={American Association for the Advancement of Science}
}

@article{bild2015aggregate,
  title={Aggregate characterization of user behavior in Twitter and analysis of the retweet graph},
  author={Bild, David R and Liu, Yue and Dick, Robert P and Mao, Z Morley and Wallach, Dan S},
  journal={ACM Transactions on Internet Technology (TOIT)},
  volume={15},
  number={1},
  pages={1--24},
  year={2015},
  publisher={ACM New York, NY, USA}
}

@article{bretherton1999effective,
  title={The effective number of spatial degrees of freedom of a time-varying field},
  author={Bretherton, Christopher S and Widmann, Martin and Dymnikov, Valentin P and Wallace, John M and Blad{\'e}, Ileana},
  journal={Journal of climate},
  volume={12},
  number={7},
  pages={1990--2009},
  year={1999}
}

@article{guess2023social,
  title={How do social media feed algorithms affect attitudes and behavior in an election campaign?},
  author={Guess, Andrew M and Malhotra, Neil and Pan, Jennifer and Barber{\'a}, Pablo and Allcott, Hunt and Brown, Taylor and Crespo-Tenorio, Adriana and Dimmery, Drew and Freelon, Deen and Gentzkow, Matthew and others},
  journal={Science},
  volume={381},
  number={6656},
  pages={398--404},
  year={2023},
  publisher={American Association for the Advancement of Science}
}

@article{larooij2026validation,
  title   = {Validation is the central challenge for generative social simulation: a critical review of {LLMs} in agent-based modeling},
  author  = {Larooij, Maik and T{\"o}rnberg, Petter},
  journal = {Artificial Intelligence Review},
  volume  = {59},
  pages   = {15},
  year    = {2026}
}

@article{ashokkumar2026large,
  title   = {Large language models can predict the results of social science experiments},
  author  = {Ashokkumar, Ashwini and Hewitt, Luke and Ghezae, Isaias and Willer, Robb},
  journal = {Nature},
  volume  = {656},
  pages   = {115--122},
  year    = {2026},
  doi     = {10.1038/s41586-026-10742-x}
}

@article{davidson2023platform,
  title={Platform-controlled social media APIs threaten open science},
  author={Davidson, Brittany I and Wischerath, Darja and Racek, Daniel and Parry, Douglas A and Godwin, Emily and Hinds, Joanne and Van Der Linden, Dirk and Roscoe, Jonathan F and Ayravainen, Laura and Cork, Alicia G},
  journal={Nature Human Behaviour},
  volume={7},
  number={12},
  pages={2054--2057},
  year={2023},
  publisher={Nature Publishing Group UK London}
}

@misc{ec2025xfine,
  author       = {{European Commission}},
  title        = {Commission fines {X} {\texteuro}120 million under the {Digital Services Act}},
  howpublished = {Press release},
  year         = {2025},
  month        = dec,
  day          = {5},
  url          = {https://digital-strategy.ec.europa.eu/en/news/commission-fines-x-eu120-million-under-digital-services-act},
  note         = {Accessed 19 October 2026}
}

@misc{australia2024minimumage,
  author       = {{Parliament of Australia}},
  title        = {Online Safety Amendment ({Social Media Minimum Age}) Act 2024},
  howpublished = {Act No.~127, 2024. Federal Register of Legislation, C2024A00127},
  year         = {2024},
  month        = dec,
  day          = {10},
  url          = {https://www.legislation.gov.au/C2024A00127/latest},
  note         = {Accessed 19 October 2026}
}

@misc{oecd2026agerestrictions,
  author       = {{OECD}},
  title        = {Social media age restrictions for children: Why they are rising and what comes next},
  howpublished = {OECD blog},
  year         = {2026},
  month        = apr,
  url          = {https://www.oecd.org/en/blogs/2026/04/social-media-age-restrictions-for-children-why-they-are-rising-and-what-comes-next.html}
}

@misc{dsa_minors_guidelines2025,
  author       = {{European Commission}},
  title        = {Communication from the Commission --- Guidelines on measures to ensure a high level of privacy, safety and security for minors online, pursuant to {Article 28(4)} of {Regulation (EU) 2022/2065}},
  howpublished = {Official Journal of the European Union, C/2025/5519},
  year         = {2025},
  month        = oct,
  day          = {10},
  url          = {https://eur-lex.europa.eu/eli/C/2025/5519/oj},
  note         = {Accessed 19 October 2026}
}

\end{document}